\documentclass[aps,prb,twocolumn,longbibliography]{revtex4-1}

\usepackage{graphicx}
\usepackage{bm}
\usepackage{braket}
\usepackage[usenames]{xcolor}
\usepackage{amsmath}
\usepackage{graphicx}
\usepackage{latexsym}
\usepackage{amsmath,amssymb}
\usepackage{xcolor}
\usepackage{stackrel}
\usepackage{accents}
\usepackage{empheq}
\usepackage[title]{appendix}
\usepackage{comment}

\usepackage{lineno}

\usepackage{float}
\usepackage{braket}
\usepackage{bbold}
\usepackage{mathtools}

\usepackage{float}
\usepackage{braket}
\usepackage{bbold}
\usepackage{mathtools}

\usepackage[normalem]{ulem}

\AtBeginDocument{%
	\newwrite\bibnotes
	\def\bibnotesext{Notes.bib}
	\immediate\openout\bibnotes=\jobname\bibnotesext
	\immediate\write\bibnotes{@CONTROL{REVTEX41Control}}
	\immediate\write\bibnotes{@CONTROL{%
			apsrev41Control,author="08",editor="1",pages="1",title="0",year="1"}}
	\if@filesw
	\immediate\write\@auxout{\string\citation{apsrev41Control}}%
	\fi
}%

\DeclareMathOperator{\Tr}{Tr}

\newcommand{\vex}[1]{\bm{\mathrm{#1}}}
\newcommand{\msf}[1]{\mathsf{#1}}

\newcommand{\w}{\omega}

\newcommand{\kb}{\vex{k}}
\newcommand{\rb}{\vex{r}}

\begin{document}

\title{Detecting Topological Superconductivity via Berry Curvature Effects in Spectral Functions}
\author{Yunxiang Liao}
\affiliation{Department of Physics, KTH Royal Institute of Technology, SE-106 91 Stockholm, Sweden}
\email{liao2@kth.se}
\author{Yi-Ting Hsu}
\affiliation{Department of Physics and Astronomy, University of Notre Dame, South Bend, IN 46556 USA}
\email{yhsu2@nd.edu}
\date{\today}

\begin{abstract}
{Experimental efforts on topological superconductivity (TSC) have primarily focused on the detection of Majorana boundary modes, while the bulk properties of TSC — particularly in two dimensions (2D) — remain relatively underexplored. 
In this work, we theoretically propose a distinctive signature in the spectral function away from the boundaries, capable of detecting 2D chiral $p$-wave TSC induced in a Rashba spin-orbit-coupled (SOC) heterostructure.
This signature can be probed experimentally through angle-resolved photoemission spectroscopy or momentum- and energy-resolved tunneling spectroscopy under a weak magnetic field $B$.
We show that within the topological phase, the spectral intensity of the lowest superconducting band at small momenta $k\sim0$ brightens (darkens) linearly with increasing $B$, whereas it darkens (brightens) in the trivial phase when the Rashba system is electron- (hole-) doped. 
This sharp contrast arises from the phase-space Berry curvature (BC) of Bogoliubov quasiparticles, a novel quantum geometric property that generalizes the conventional momentum-space BC.  
The effect of this phase-space BC can also be detected by the differential conductance away from the boundaries. 
Our falsifiable prediction provides an experimental avenue for detecting Rashba-induced chiral $p$-wave TSC without relying on Majorana mode detection, addressing a key challenge in the realization of 2D TSC.}
\end{abstract}

\maketitle

\section{Introduction}

Topological superconductors (TSCs) 
are exotic quantum materials that exhibit non-trivial band topology in the bulk as well as Majorana modes on different dimensional boundaries 
\cite{ReadGreen,Kitaev2001aa,JiaoNature2020,Khalaf2018,ZhangPRL2019,HsuPRL2020,MoTe2_Hsu}. 
The experimental detection of a TSC has remained one of the key challenges in the fields of unconventional superconductivity and topological phases, despite intensive studies in the past decade. 
Currently, many TSC experiments have focused on detecting Majorana boundary signatures through local scanning probes \cite{Feldman:2017aa,Exp_2DTscProx_Ncomm2017,Exp_MajoFeSc_Peng2018,Exp_MajoFeSc_Wang2018,Exp_MajoSTM2DTsc_Yazdani,Exp_STMMajoedge_2019,JiaoNature2020,Exp_gold_Manna2020,Exp_VdW_Kezilebieke2020,Exp_FeSc_MajoedgeSTM_2020,STMreview_Majo,Exp_Nonmajo_Frolov}, although this approach often faces the challenge of distinguishing signals originating from Majorana modes and from other more mundane sources \cite{Nonmajo_VonOppen_2015,MajoSpin_Bernevig2018,RetractMajo2018,Thy_ZBP_Sankar,Review_nonmajo_Sau,Thy_nonMajo_Sankar_PRB2020,STMreview_Majo,Exp_Nonmajo_Frolov,Exp_nonMajo_Katsaros2021}. 
In contrast, experimental features in the bulk of TSC associated with superconducting band topology remain relatively underexplored. 

Bands with non-trivial topology necessarily possess non-trivial band geometry, which characterizes the changes in eigenstates at neighboring momenta $\textbf{k}$ and could have significant influences on experimental observables in metals, insulators, and superconductors \cite{Niu1999,NiuDOS,Shindou,NiuRev,GaoPRB2015,JulkuPRL2016,LiangPRB2017,LiangBCPRB2017,
	NiuSC,RossiReivew2021,AhnPRB2021,YuQG2023}.  
Though relatively unexplored, quantum geometric quantities defined in real-space and phase-space can also lead to detectable experimental features.
For example, non-trivial Berry curvature (BC) $\Omega_{\lambda\lambda'}, \lambda, \lambda' =r,k$, is expected for the Bogoliubov quasiparticles in two-dimensional (2D) chiral TSC, which features a non-zero Chern number $C=\int d\textbf{k}\Omega_{kk}(\textbf{k})$
and quantized
thermal Hall conductivity\cite{ReadGreen}. 
Besides the momentum-space BC, 
the generalized BCs defined in the real space $\Omega_{rr}$ and the phase space $\Omega_{kr}$ generally exist in the presence of external perturbations, such as a magnetic field or supercurrent \cite{NiuSC}.
These generalized BCs were recently found to influence the density of states and thermal responses in spin-singlet superconductors with spin-degenerate normal states \cite{NiuSC}. 

\begin{figure}[t]
	\centering
	\includegraphics[width=8cm]{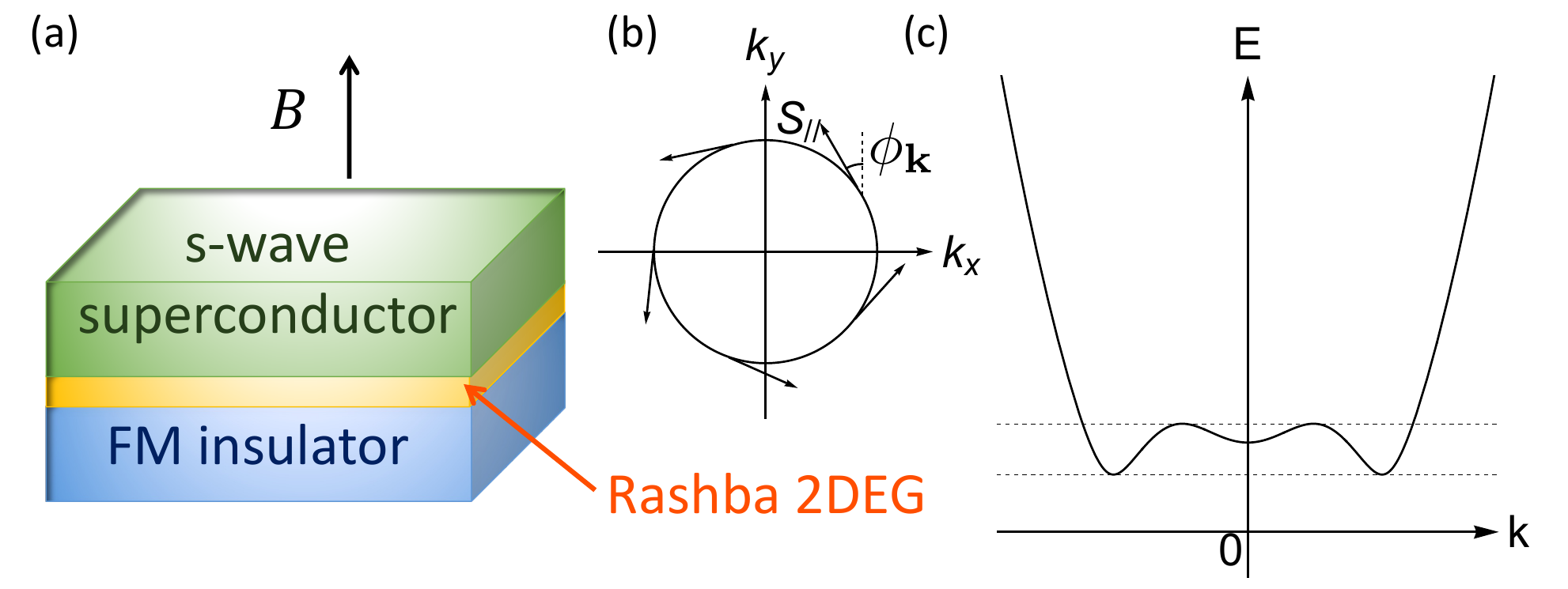}
	\caption{
		Sketches for (a) the considered heterostructure hosting an effective 2D TSC, and (b) the in-plane spin texture $S_{\parallel}$ on the Fermi surface of the Rashba layer, where $\phi_\textbf{k}$ labels the intrinsic locking angle between the spin and momentum.  (c) The calculated lower BdG band $E_{-}(k)$ of $H$, where the dotted lines label the band extrema.
	}
	\label{fig:0}
\end{figure}

Spin-orbit-coupled (SOC) normal states, on the other hand, have been a crucial ingredient in strategies to achieve  TSC since manipulating Fermi surface spin textures can facilitate the formation of effectively spin-triplet Cooper pairs
\cite{FuKane_TITsc,RashbaTsc_Lee2009,SauPRL2010,Sau,Brouwer2011,ZhangPRB2011,SOCTsc_PotterLee_2012,SOCchiralp_OjanenPRL2015,Li:2016aa,Exp_2DTscProx_Ncomm2017,Hsu2017,Exp_gold_Manna2020,Exp_VdW_Kezilebieke2020,Wang_NanoLett2022}. 
Candidate materials for 2D chiral TSC predicted from this strategy include the 
proximitized surfaces of topological insulators~\cite{FuKane_TITsc} and hole-doped monolayer transition metal dichalcogenides \cite{Hsu2017}.  
{In this work, we will focus on proximitized two-dimensional gas (2DEG) with Rashba SOC~\cite{Sau} (see Fig. \ref{fig:0}a).}


Here, we theoretically predict a BC-induced signature in the spectral function that distinguishes whether the superconducting Rashba 2DEG in Fig. \ref{fig:0}a is in the topological or trivial phase.
This signature is found by examining the generalized BC effects of Bogoliubov quasiparticles using a semiclassical wavepacket analysis \cite{NiuRev,NiuSC}, and can be probed by angular-resolved photoemission spectroscopy (ARPES) or momentum- and energy-resolved tunneling spectroscopy (MERTS)\cite{MERTS} under a weak magnetic field $B$\cite{ARPES_field}. 
We show that the intensity of the lowest superconducting band at small momenta $k\sim0$ brightens (darkens) linearly with $B$ in the topological phase but darkens (brightens) in the trivial phase, when the Rashba 2DEG is electron-(hole-)doped.  
This phase-space BC effect can therefore serve as an experimental diagnostic for chiral $p$-wave TSC realized by a proximitized Rashba 2DEG.


\section{Model}

 We consider a well-known minimal model \cite{SauPRL2010,Sau} that describes the chiral $p$-wave TSC resulting from a Rashba 2DEG proximitized by a $s$-wave superconductor and a ferromagnetic insulator (see Fig.~\ref{fig:0}). 
In the presence of a weak magnetic field $\vex{B}=B \hat{z}$, the 2DEG layer is described by 
\begin{align}\label{eq:H-0}
\begin{aligned}
H=&
\sum_{\sigma=\uparrow,\downarrow}
\int d^2\rb
c_{\sigma}^{\dagger}(\rb) 
\left( 
\frac{1}{2m}\left( -i\nabla-\vex{A}(\rb)\right) ^2-\mu
+h\sigma_z
\right) 
c_{\sigma}(\rb)
\\
&+
\sum_{\sigma=\uparrow,\downarrow}
\int d^2\rb
c_{\sigma}^{\dagger}(\rb) 
\alpha (\vex{\sigma} \times \left( -i\nabla-\vex{A}(\rb) \right) )\cdot \hat{z}
c_{\sigma}(\rb)
\\
&+
\int d^2\rb
\left( 
\Delta
c^{\dagger}_{\uparrow}(\rb)
c^{\dagger}_{\downarrow}(\rb)
+
\Delta^*
c_{\downarrow}(\rb)
c_{\uparrow}(\rb)
\right),
\end{aligned}
\end{align}
where $c_{\sigma}^{\dagger}(\rb)$ creates an electron at position $\rb$ with spin $\sigma=\uparrow,\downarrow$ and mass $m$, ${\sigma}_{i}$ are spin Pauli matrices, 
and $\vex{A}(\rb)=\vex{B} \times \rb/2$ denotes the vector potential. 
We have set 
$\hbar=1$, the electric charge $e=1$, and $h>|\mu|$ so that the chemical potential $\mu$ intersects only the lower normal band. 
The second and third lines describe respectively the Rashba SOC with strength $\alpha$ and the proximity-induced superconductivity with a pairing potential $\Delta$. 

{This model is in the chiral $p$-wave TSC phase when the ferromagnetic exchange coupling $h>\sqrt{|\Delta|^2+\mu^2}$ and in the trivial phase when $h<\sqrt{|\Delta|^2+\mu^2}$. In the following calculations, we consider a weak magnetic field $B<|\Delta|$ and a Rashba layer within the TSC or trivial phase. For simplicity, we approximate $h$ and $B$ as two independent variables, assuming that the Zeeman splitting caused by the applied magnetic field and the orbital effect of the ferromagnetic insulator are negligible. 

We examine the dynamics of quasiparticles described by $H$ using a semiclassical wavepacket approach, which has been successfully applied to metals \cite{Niu1999,NiuRev,Niu2020} 
and spin-singlet superconductors with spin-degenerate normal states  \cite{Spivak,LiangPRB2017,NiuSC}. 
We assume the external perturbation $\vex{A}(\rb)$ is time independent and slowly varying in the real space \vex{r} compared to the spread of the wavepacket.
This allows us to neglect the \vex{r} dependence in $\vex{A}(\rb)$ in Eq.~\ref{eq:H-0} and approximating $\vex{A}(\rb)$ by its value at the wavepacket center $\rb=\rb_c$. 
Such a scenario can be achieved when the applied field strength is smaller but close to $H_{c2}$ so that the lattice constant of the vortex lattice is close to the coherence length. 


The resulting effective Hamiltonian governing the wavepacket dynamics can be further simplified by introducing the  Nambu spinor 
	\begin{align}
	\Psi(\kb)
	=
	\begin{bmatrix}
	c_{\uparrow}(\kb)
	&
	c_{\downarrow}(\kb)
	&
	c^{\dagger}_{\downarrow}(-\kb)
	&
	-c^{\dagger}_{\uparrow}(-\kb)
	\end{bmatrix}^{T},
	\end{align}
    where 
	\begin{align}\label{eq:FT}
	\begin{aligned}
		c_{\sigma}^{\dagger}(\kb)
		=\,&
		\int
		d^2\rb
		e^{i \left( \kb+ \vex{A}(\rb_c)\right) \cdot \rb}
		c_{\sigma}^{\dagger}(\rb),
	\end{aligned}
	\end{align}
	After the transformation, the Hamiltonian assumes the form
			\begin{align}\label{eq:H-1}
	\begin{aligned}
	H=\,&
	\frac{1}{2}
	\sum_{\sigma=\uparrow,\downarrow}
	\int \frac{d^2\kb}{(2\pi)^2}
	\Psi^{\dagger}(\kb) h_{\msf{BdG}} (\kb)\Psi(\kb),
	\\
	h_{\msf{BdG}}(\kb)=\,&
	\left[ \xi_k+\alpha (\sigma_x k_y-\sigma_y k_x) \right]  \tau_{z}+h\sigma_z
    \\
    &
	+\frac{1}{2}\Delta(\tau_{x}+i\tau_{y})+\frac{1}{2}\Delta^*(\tau_{x}-i\tau_{y}).
	\end{aligned}
	\end{align}
	Here $\xi_{k}= k^2/2m-\mu$ and $\tau_{i}$ denotes the Pauli matrix acting in the Nambu space.

This Hamiltonian can be diagonalized to
\begin{align}\label{eq:H-2}
\begin{aligned}
    H=\,&
    \sum_{s=\pm 1}
    \int \frac{d^2\kb}{(2\pi)^2}
    E_{s}(k) \gamma^{\dagger}_s(\kb)\gamma_s(\kb)
    +\text{const.},
\end{aligned}
\end{align}
by applying the Bogoliubov–de Gennes (BdG) transformation:
	\begin{align}\label{eq:BdG}
	\begin{aligned}
		&\gamma_{s}(\kb)
		=\,
		\Phi^{\dagger}_{s}(\kb)
		\Psi(\kb),
	\end{aligned}
\end{align}
where the helicity $s=\pm$ labels the upper and lower BdG bands.
$\Phi_{s}(\kb)
=
\begin{bmatrix}
u_{\uparrow s}(\kb)
&
u_{\downarrow  s}(\kb)
&
v_{\downarrow s}(\kb)
&
v_{\uparrow  s}(\kb)
\end{bmatrix}^{T}$   is the eigenfunction of $h_{\msf{BdG}}(\kb)$ with positive eigenenergy $E_{s}(k)$ and it satisfies
\begin{align}\label{eq:BdGEq}
\begin{aligned}
h_{\msf{BdG}}(\kb)
\Phi_{s}(\kb)
=
E_{s}(k)
\Phi_{s}(\kb).
\end{aligned}
\end{align}

Solving the  BdG equation, we find the quasiparticle energy 
\begin{align}
\begin{aligned}
E_{s}(k)=&\,
\left[(\alpha k)^2+|\Delta|^2+h^2+\xi_k^2
\right.
\\
&
\left.
+2 s \sqrt{\xi_k^2 \left((\alpha k)^2+h^2\right)+|\Delta|^2 h^2}
\right]^{1/2}.
\end{aligned}
\end{align}
For $\xi_k\neq h$ or $s\neq -1$, the BdG eigenfunction 
$\Phi_{s}(\kb)$ 
is given by,
\begin{align}\label{eq:coh}
\begin{aligned}
u_{\uparrow  s}(\kb)
=\,&
D_{s}(k)
\left[ E_{s}(k) h\xi_{k}
+|\Delta|^2 h+h \xi_{k} ^2+ \xi_{k} (\alpha^2k^2+h^2)
\right.
\\
&\left.
+ s (E_{s}(k) +  h+ \xi_{k})\sqrt{(\alpha k)^2 \xi_k^2+h^2 \left(|\Delta|^2+\xi_k^2\right)}
 \right]
 \\
 &\times
e^{i (\chi-\phi_{\kb}+\pi/2)/2},
\\
u_{\downarrow  s}(\kb)
=\,&
D_{s}(k)
\alpha k 
\left[
\xi_{k}  (\xi_{k} +E_{s}(k) )
\right.
\\
&\left.
+ s \sqrt{(\alpha k)^2 \xi_k^2+h^2 \left(|\Delta|^2+\xi_k^2\right)}
\right]
e^{i (\chi+\phi_{\kb}-\pi/2)/2},
\\
v_{ \downarrow  s}(\kb)
=\,&
D_{s}(k)
|\Delta| 
\left[
h (h+ E_{s}(k) )
\right.
\\
&\left.
+ s \sqrt{(\alpha k)^2 \xi_k^2+h^2 \left(|\Delta|^2+\xi_k^2\right)}\right]
e^{- i (\chi+\phi_{\kb}-\pi/2)/2},
\\
v_{\uparrow  s}(\kb)
=\,&
D_{s}(k)
|\Delta|\alpha k   (\xi_{k}-h)
e^{- i (\chi-\phi_{\kb}+\pi/2)/2}.
\end{aligned}
\end{align}
By contrast, when $\xi_{k}=h$ and $s=-1$, we instead have
\begin{align}\label{eq:coh-2}
\begin{aligned}
u_{\uparrow ,-1}(\kb)
=\,&
0,
\\
u_{\downarrow , -1}(\kb)
=\,&
D_{-1}(k)
|\Delta|
e^{i (\chi+\phi_{\kb}-\pi/2)/2},
\\
v_{ \downarrow , -1}(\kb)
=\,&
-D_{-1}(k)
\alpha k
e^{- i (\chi+\phi_{\kb}-\pi/2)/2},
\\
v_{\uparrow,  -1}(\kb)
=\,&
D_{-1}(k)
E_{-1} (k)
e^{- i (\chi-\phi_{\kb}+\pi/2)/2}.
\end{aligned}
\end{align}
Here 
the phase of the pairing potential $ \chi\equiv\arg(\Delta)$ depends on the pairing symmetry~\cite{FN0},
the intrinsic angle of the Rashba SOC is denoted by $ \phi_{\kb}\equiv \arctan(k_y/k_x)$.
$D_{s}(k)$ is the normalization constant.


\section{Quasiparticle wavepacket}\label{sec:wave}

After diagonalizing $H$, we 
construct a wavepacket for the  quasiparticles from the lower BdG band with $s=-$: 
\begin{align}
\begin{aligned}
\ket{W_s}
=\,
\int_{\kb} w_s(\kb,t)
\gamma_{s}^{\dagger}(\kb)
\ket{G}, 
\end{aligned}
\end{align}
where $\ket{G}$ denotes the superconducting ground state. 
The envelope function $w_s(\kb,t)$ 
obeys the normalization condition
$
\int_{\kb} |w_s(\kb,t)|^2=1 
$
and is sharply peaked around $\kb_c=
\int_{\kb} |w_s(\kb,t)|^2\kb$, 
whereas the wavepacket center in the real space is given by
~\cite{LiangPRB2017} 
\begin{align}\label{eq:rc}
\begin{aligned}
 \rb_c
=\, &
i
\int_{\kb}
\left( 
w_s(\kb,t)
\Phi_s(\kb)
\right)^{\dagger} 
\partial_{\kb}
\left( 
w_s(\kb,t)
\Phi_{s}(\kb)
\right) 
.
\end{aligned}
\end{align}
For cases where the perturbation is strong, a treatment involving both $s=\pm$ bands is required \cite{NiuMult,Shindou,NiuRev}, which falls beyond the scope of our current study.

The semiclassical dynamics of this quasiparticle wavepacket $\ket{W_s}$ 
is described by 
its Lagrangian $L$,
given by the wavepacket average~\cite{FN1} of the operator 
$\hat{L}= i \frac{d}{dt} - \hat{H}$~\cite{NiuRev,NiuSC}. It
can be expressed in terms of the momentum-space and real-space Berry connections $\mathcal{A}_{ks}$ and ${\mathcal{A}}_{rs}$ as
\begin{align}\label{eq:L}
\begin{aligned}
L
=\,	&
-
E_{s}(k_c)
-
\dot{\kb}_c
\cdot
\rb_c
+
\dot{\kb}_c
\cdot
\mathcal{A}_{ks}
+
\dot{\rb}_c 
\cdot 
{\mathcal{A}}_{rs}.
\end{aligned}
\end{align}
The momentum-space Berry connection $\mathcal{A}_{ks}$ is defined in terms of the BdG wavefunction $\Phi_{s}(\kb)$ as $\mathcal{A}_{ks} 
=\, 
i\Phi^{\dagger}_s(\kb_c)
\partial_{\kb_c} \Phi_s(\kb_c)$, 
and can be further simplified into
\begin{align}\label{eq:Aks-1}
\begin{aligned}
\mathcal{A}_{ks} 
=&
-\frac{1}{2}\rho_{0s} (\kb_c) \partial_{\kb_c}  \chi
+\frac{1}{2}\rho_{1s} (\kb_c) \partial_{\kb_c}  \phi_{\kb_c}. 
\end{aligned}
\end{align}
Here
\begin{align}\label{eq:rho}
\rho_{a s}(\kb)
=
\sum_{\sigma}
(\zeta_{\sigma})^{a}
\left( 
|u_{\sigma s}(\kb)|^2
-
|v_{\sigma s}(\kb)|^2 
\right)
\end{align}
 with $a=0,1$ and $\zeta_{\uparrow/\downarrow}=\pm 1$.
Physically speaking, $\rho_{0s}(\kb_c)$ and $\rho_{1s}(\kb_c)/2$ correspond to the wavepacket average~\cite{FN1} of the total charge 
and spin,  
respectively (see Appendix \ref{app:wave}). 

The real-space Berry connection ${\mathcal{A}}_{rs}$ 
is given by
\begin{align}\label{eq:tArs}
\begin{aligned}
{\mathcal{A}}_{rs}
= &
\tilde{\mathcal{A}}_{rs}
+
\rho_{0 s}(\kb_c) \vex{A}(\rb_c)
+
\frac{1}{2}
\vex{B} \times \vex{d}, 
\end{aligned}
\end{align}
\text
where $\tilde{\mathcal{A}}_{rs}$ is the real-space analogue of  $\mathcal{A}_{ks}$ in Eq.~\ref{eq:Aks-1} by replacing   $\partial_{\kb_c}\rightarrow \partial_{\rb_c}$.
The second term describes how the total charge of the wavepacket $\rho_{0s}(\kb_c)$ couples to the external vector potential $\vex{A}(\rb_c)$, whereas  
the last term describes how the charge dipole moment \textbf{d} of the wavepacket couples to the external magnetic field $\vex{B}$. 
Note that these two terms enter the Lagrangian in Eq. \ref{eq:L} in the same way as the well-known Lagrangian describing the Lorentz force of a charged object with a dipole moment in a magnetic field \cite{LL}. 

The dipole moment \textbf{d} of the wavepacket, given by the wavepacket average~\cite{FN1} of dipole moment operator 
$\hat{\vex{d}}
=\,
\sum_{\sigma}
\int d^2 \rb \,
\hat{c}_{\sigma }^{\dagger}(\rb)
\hat{c}_{\sigma }(\rb)
(\rb-\rb_c)
$, contains two terms $\vex{d}=\vex{d}_1+\vex{d}_2$. The first term  
\begin{align}\label{eq:d1}
\vex{d}_1
=\,
\frac{1}{2}
\left( \rho_{0s}^2(\kb_c)-1\right) 
\partial_{\kb_c} \chi,
\end{align}
was found nonzero for a 
$d$-wave pairing potential \cite{NiuSC}, but vanishes for our case of an $s$-wave pairing potential with $\partial_{\kb} \chi=0$.
Nonetheless, we find that the second term 
\begin{align}\label{eq:P-0}
\begin{aligned}
\vex{d}_2
=\,&
\frac{1}{2}
\left( \rho_{2 s}(\kb_c)-\rho_{0s}(\kb_c)\rho_{1s} (\kb_c)\right) 
\partial_{\kb_c} \phi_{\kb_c}   ,
\end{aligned}
\end{align}    
is a new source of dipole moment arising from the SOC. Note that although $\textbf{d}_2$ scales with the winding of the Rashba angle $\phi_{\kb_c}$ (Fig. 1), it is \textit{not} inherited from the normal state because the prefactor $\rho_{2 s}-\rho_{0 s}\rho_{1 s}=0$ when the pairing potential $\Delta=0$, where  \begin{align}\label{eq:rho2}
\rho_{2 s}(\kb_c)=\sum_{\sigma}\zeta_{\sigma}\left( |u_{\sigma s}(\kb_c)|^2+|v_{\sigma s}(\kb_c)|^2\right).
\end{align}
Thus, $\textbf{d}_2$ exists only in the superconducting state but not the normal state. 
This is physically because the quasiparticle wavepacket consists of momentum- and position-dependent mixtures of electron and hole components, which allows a deviation of the wavepacket center $\textbf{r}_c$ from the charge center.
Since the dipole moment $\vex{d}_2$ emerges exclusively when superconductivity and spin-orbit coupling coexist, $\vex{d}_2$ can lead to new experimental features of SOC superconductors through coupling to an external  $\textbf{B}$ field (see Eq. 6).

\section{Phase-space BC and density of states}

To analyze the experimental features of $\vex{d}_2$-induced BC, 
we examine how the wavepacket dynamics are modified by BC from Eq.~\ref{eq:L}:  
 \begin{align}\label{eq:EOM-main}
 \begin{aligned}
	&\begin{bmatrix}
	\Omega_{rrs}^{ij}  
	&
	\Omega_{rks}^{ij}-\delta_{ij}
	\\
	\Omega_{krs}^{ij}+\delta_{ij}
	&
	\Omega_{kks}^{ij}
	\end{bmatrix}
	\begin{bmatrix}
	\dot{\rb}_c^{j}
	\\
	\dot{\kb}_c^{j}
	\end{bmatrix}
	=
	\begin{bmatrix}
	\partial_{r_c^{i}} E_{s}(k_c)
	\\
	\partial_{k_c^{i}} E_{s}(k_c)
	\end{bmatrix}, 
 \end{aligned}
 \end{align} 
where the repeated Cartesian indices are implicitly summed. 
The momentum-, real-, and phase-space BCs 
are defined as the derivatives of the Berry connections ${\mathcal{A}}_{rs}$ and  $\mathcal{A}_{ks}$: 
\begin{align}
\Omega_{\lambda\lambda's}^{ij}
\equiv\,
\partial_{\lambda^{i}_c} {\mathcal{A}}_{\lambda's}^{j}
-\partial_{\lambda'^{j}_c} {\mathcal{A}}_{\lambda s}^{i},~
\lambda, \lambda' = r, k.
\end{align}
For the $p+ip$ TSC we consider in Eq.~\ref{eq:H-0}, we find that the momentum- and phase-space BC under an external \textbf{B} field are given by (see Appendix~\ref{app:EOM})
\begin{align}\label{eq:vectorOmega}
\begin{aligned}
{\bf \Omega}_{ks}
=\,&
\frac{1}{2}\partial_{\kb_c} \rho_{1s}(\kb_c)  \times \partial_{\kb_c}  \phi_{\kb_c},
~\Omega_{rks}^{ij}
	= \,
    -
    \frac{1}{2}\partial_{k^{j}_c} (\vex{B}\times \vex{d})^{i} ,
\end{aligned}
\end{align} 
up to the leading order in the real-space gradient expansion, 
where field-induced supercurrent is not considered in this work. 
We find that while the presence of real-space BC ${\bf \Omega}_{rs}
=\,
\rho_{0 s}(\kb_c)\vex{B}$ 
does not depend on SOC, 
neither ${\bf \Omega}_{ks}$ nor $\Omega_{rks}^{ij}$ would have survived in the absence of the SOC. Specifically, the momentum-space ${\bf \Omega}_{ks}$ is inherited  from the spin-orbit coupled normal state, whereas the phase-space $\Omega_{krs}$ describes the coupling between the SOC-induced wavepacket dipole moment to the external \vex{B} field. 




The BC effects on the semiclassical dynamics in Eq.~\ref{eq:EOM-main} have been shown~\cite{NiuDOS,NiuRev} to modify the phase-space density of states $D(\rb,\kb)$, required by the conservation of the total number of states $\mathcal{N}= D(\rb,\kb)\Delta V$ in a phase-space volume $\Delta V=\Delta k \Delta r$ (see Appendix~\ref{app:DOS}). 
The modified phase-space density of states  was shown to be~\cite{NiuRev,Arnold,NiuSC} 
$D(\rb,\kb)\approx 1+
\Tr \Omega_{krs}
-
{\bf \Omega}_{rs} \cdot {\bf \Omega}_{ks}  
$ 
up to an inessential overall coefficient, 
where a spatial gradient expansion is applied for slowly varying external fields. 
This BC-modified phase-space density of states $D(\rb,\kb)$ is the key factor that embeds the information of the wavepacket BCs we find in Eq.~\ref{eq:vectorOmega} into experimental observables. 

For the $p+ip$ TSC we consider, we obtain  
the BC-induced phase-space density of states $\delta D=D-1$ under an external \textbf{B} field using  Eq.~\ref{eq:vectorOmega} 
\begin{align}\label{eq:dD1}
\delta D
=\,
		-
		\frac{1}{2}
		\vex{B}  \cdot (\nabla_{\kb} \times \vex{d}_2 )
		-
		\frac{1}{2}
(
		\rho_{0s} 
		\nabla_{\kb} \rho_{1s}  \times \nabla_{\kb}   \phi_{\kb}
  )
		\cdot
		\vex{B} .
\end{align}
up to the leading order in spatial gradient $\partial_{\rb}$.  
Note that $\delta D$ exists only in the presence of SOC: 
The latter term results from the momentum- and real-space BC, which is proportional to the winding of the Rashba angle $\phi_{\kb}$ in the spin-orbit-coupled normal state.  
The former term 
results from 
the phase-space BC, which exists only in a spin-orbit-coupled superconductor due to  
the quasiparticle charge dipole moment $\vex{d}_2$ in Eq. 7.   

\begin{figure}[t!]
	\centering
	\includegraphics[width=8.2cm]{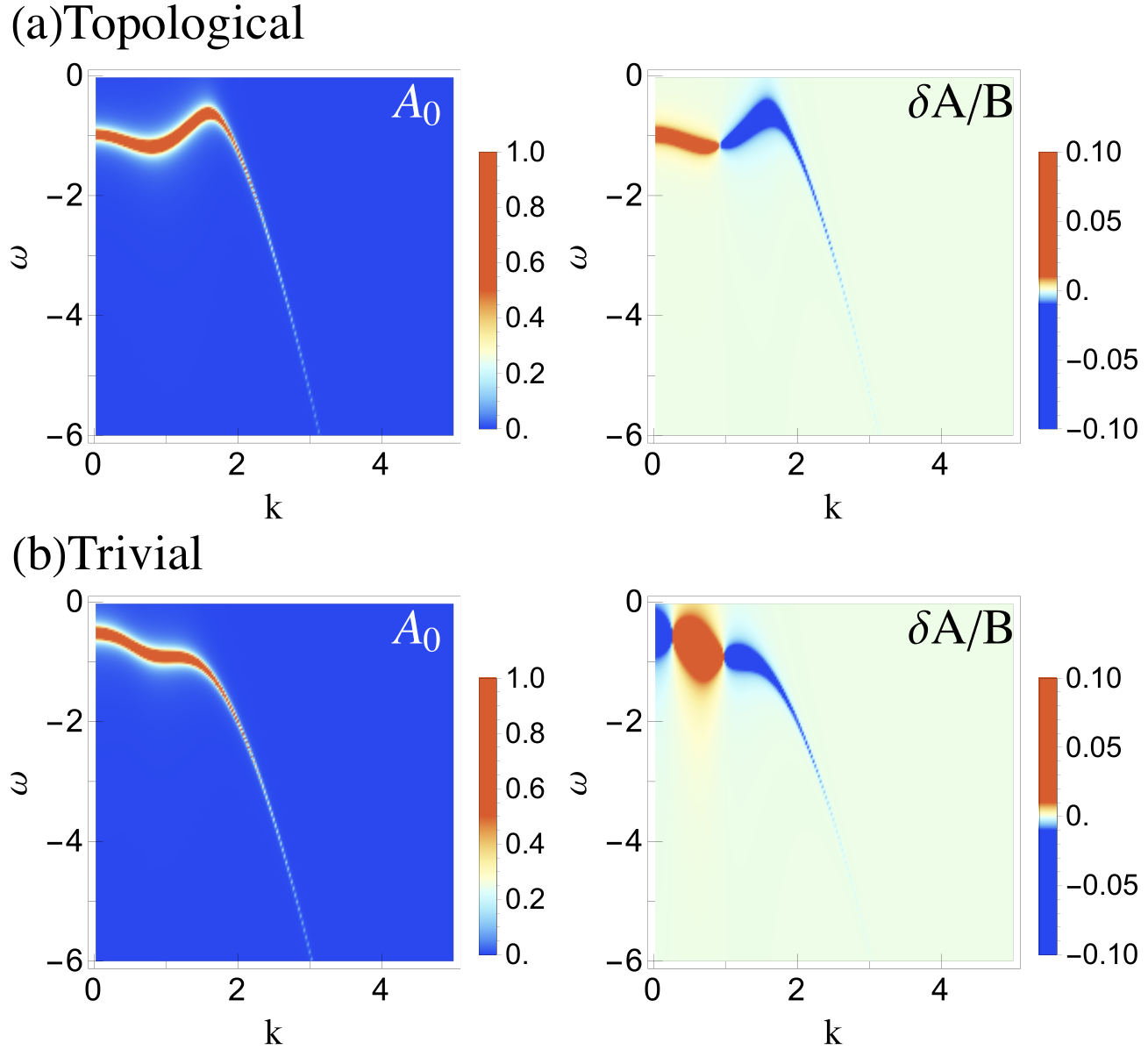}
 	\caption{
		BC effects on the the spectral function $A(\textbf{k},\omega)$ in the (a) topological phase and (b) trivial phase. The bare spectral function $A_0(\kb,\omega)$ and
 the BC contribution per unit field strength $\delta A(\kb,\omega)/B$ are shown on the left and right panels, respectively. Here, we set $h=2$ and $h=0.5$ in (a) and (b), respectively, whereas we fix the pairing potential $\Delta=1$, chemical potential $\mu=0.2$, mass $m=0.5$, and SOC strength $\alpha=1$ in all figures unless specified otherwise. This figure and Fig.~\ref{fig:1} are generated by replacing the Dirac delta function in Eq.~\ref{eq:NI} with a Lorentzian $L(\w)=\frac{1}{\pi} \frac{\eta/2}{\w^2+(\eta/2)^2}$ of width $\eta=0.05$. 
} 
\label{fig:2}
\end{figure}

\section{BC-induced Topological Signatures in Observables}

We are now ready to examine how the BCs influence the experimental observables in a $p+ip$ TSC enabled by Rashba SOC. 
With the BC-modified 
phase-space density of states $D=1+\delta D$, we investigate the spectral function $A(\kb,\omega)$ and the normal metal-to-superconductor tunneling conductance~\cite{Schrieffer}  $G(\textbf{r},\omega)$
given by  
\begin{subequations}\label{eq:NI}
    \begin{align}
  &\begin{aligned}\label{eq:I}
	A(\kb,\omega)
	=\,& 
	\sum_{\sigma}
        \int_{\rb}
	D(\rb,\kb)
        	\left[ 
	 |u_{\sigma s}(\kb)|^2
	\delta \left( \omega-E_{s}(k) \right) 
	\right. 
	\\
	&
	\left. 
	+
	 |v_{\sigma s}(\kb)|^2
	\delta \left( \omega+E_{s}(k) \right) 
	\right],  
 	\end{aligned}
	\\    
   &\begin{aligned}\label{eq:N}
	G(\rb,\omega)
	=\,&
	\sum_{\sigma}
	\int_{\kb} 
	D(\rb,\kb)
	\left[ 
	 |u_{\sigma s}(\kb)|^2
	\delta \left( \omega-E_{s}(k) \right) 
	\right. 
	\\
	&
	\left. 
	+
	 |v_{\sigma s}(\kb)|^2
	\delta \left( \omega+E_{s}(k) \right) 
	\right],
	\end{aligned}	
	\end{align}
\end{subequations} 
where we focus on the lower BdG band $s=-$ and neglect the small spatial variations in $D(\rb,\kb)$ given the slowly varying perturbation. 
The spectral function can be probed by the momentum- and energy-resolved tunneling spectroscopy MERTS\cite{MERTS} or in principle ARPES under a weak magnetic field. 
The tunneling conductance can be measured by STM or tunneling junctions. For STM measurements that require assess to open surfaces, a bilayer structure\cite{Exp_VdW_Kezilebieke2020} could be more feasible than the heterostructure in Fig.~\ref{fig:0}(a).

First, we demonstrate the BC effects in the spectral function $A=A_0+\delta A$ in Fig. \ref{fig:2}. 
The uncorrected spectral function $A_0(\kb,\omega)$ and the BC contribution $\delta A(\kb,\omega)$ are obtained by setting $D(\kb)=1$ and $\delta D(\kb)$ in Eq.~\ref{eq:I}, respectively. 
At zero field $\vex{B}=0$, the magnitude of $A_0$ reaches its peaks at frequencies along the lower superconducting band $\omega=-E_{-}(\kb)$ (see Fig.~\ref{fig:2} (a)). 
Under an applied field $B$, the BC contribution $\delta A$ is activated, which modulates the peak intensities in $A_0$ in a momentum-dependent way tied to topology:   
In the topological phase where $h>\sqrt{\Delta^2+\mu^2}$, 
the peak intensity at small momenta $\textbf{k}\sim 0$ is enhanced by $\delta A>0$,  as shown in Fig.~\ref{fig:2}(a). In contrast, in the topologically trivial phase where $h<\sqrt{\Delta^2+\mu^2}$, we find a suppression $\delta A<0$ at $\textbf{k}\sim 0$, as shown in Fig.~\ref{fig:2}(b). 
In either phase, the magnitude of BC-induced enhancement or suppression $|\delta A|$ increases linearly with $B$ until it reaches the critical field strength.

Now, we show how the topological conditions explicitly enter the BC-modified spectral function $A=A_0+\delta A$ at $\textbf{k}=0$. 
For superconductivity described by the model $H$ in Eq. \ref{eq:H-0}, the topological conditions refer to (1) the Rashba strength $\alpha\neq 0$ is nonzero and (2) $h^2 >(\Delta^2 + \mu^2)$. 
To link these two conditions to
the BC effect in the spectral function, we insert the full expressions of the coherence factors $u_{\sigma s}(\textbf{k}_c)$ and $v_{\sigma s}(\textbf{k}_c)$ into the BC-induced change $\delta D(\textbf{k})$ in the phase-space density of states (see Eq. \ref{eq:dD1}) and arrive at
\begin{align}\label{eq:deltaDk0}
\begin{aligned}
    \delta D(k=0)
    =
    B\alpha^2 
    \dfrac{ \mu(h\Delta^2+(h^2-\mu^2)\sqrt{\Delta^2+\mu^2})}{2h^2(\Delta^2+\mu^2)(h^2-\Delta^2-\mu^2)}. 
\end{aligned}
\end{align}
It is clear from Eq. \ref{eq:deltaDk0} that when only the lower normal band involves in pairing $h>|\mu|$, criterion (1) and (2) indicate that the BC-induced change
{
$\delta D(k=0)$ 
takes the sign of the chemical potential $\mu$ in the topological phase and its opposite in the trivial phase, respectively}. 
Together with the fact that the normalized BC contribution $\delta A(\textbf{k},\omega)/A_0(\textbf{k},\omega)=\delta D(\textbf{k},\omega)$ at 
$\omega=-E_{-}(\kb)$, which follows from Eq. \ref{eq:I}, we conclude that the sign of the BC effect $\delta A$ at momentum 
$\textbf{k}=0$ qualitatively detects whether the superconductivity is in the topological or trivial phases.  

Experimentally, this topological signature can be {detected in the following way: First,  determine whether the chemical potential of the 2DEG is $\mu>0$ or $\mu<0$, i.e. whether $\mu$ is closer to the higher or lower band bottom. Then in the former (latter) case, one can identify whether the 2D superconductivity is in the topological (trivial) or trivial (topological)  phase if the ARPES or MERTS signal intensity enhances $\delta A>0$ or suppresses $\delta A\leq 0$ when an applied weak magnetic field is applied. This change in intensity $\delta A$ is expected to grow linearly with the field strength $B$ \cite{FN2}.} 


\begin{figure}[t!]
\centering
\includegraphics[width=1\linewidth]{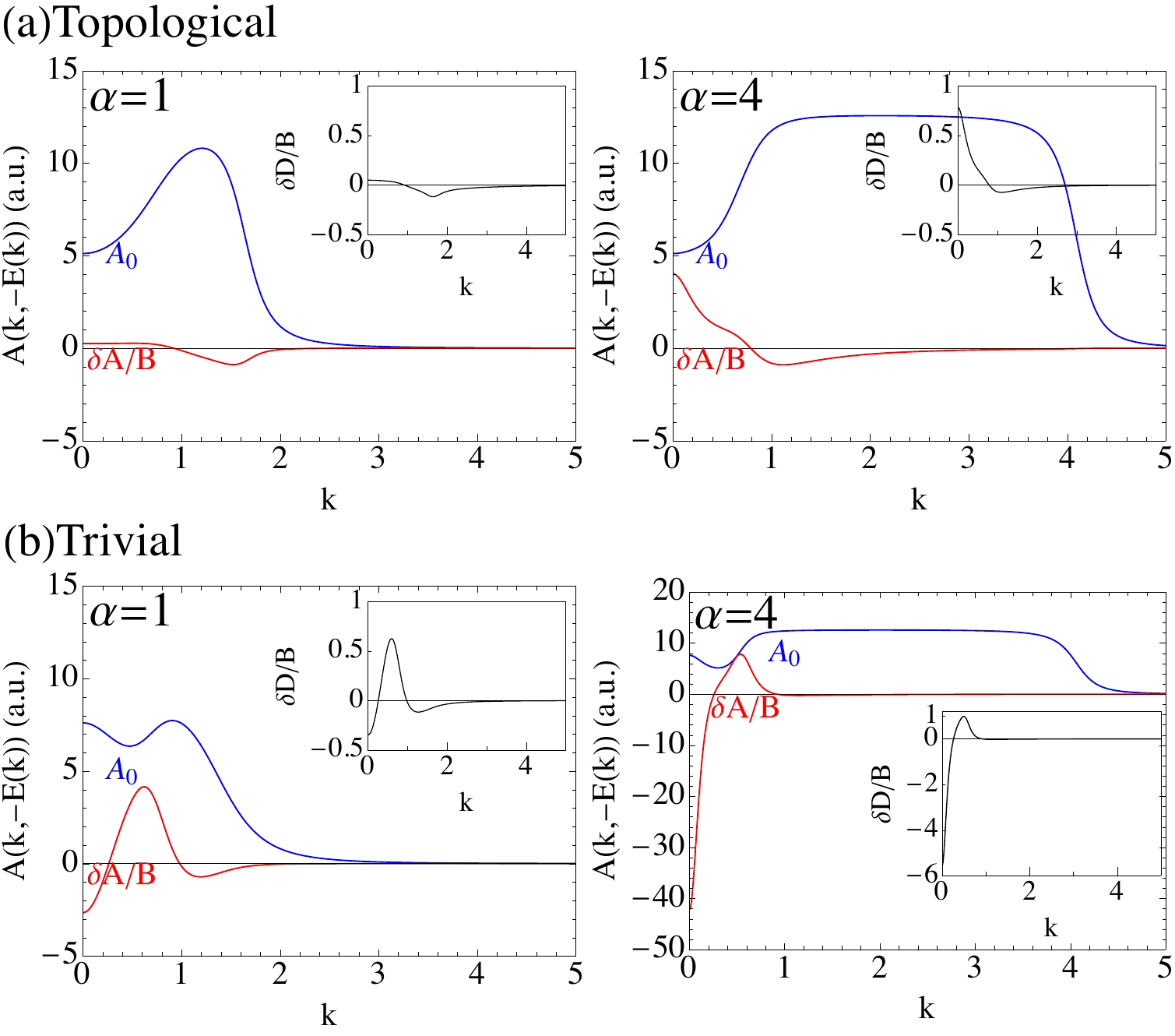}
\caption{
{The bare spectral functions $A_0(\kb,\omega)$ and BC contribution per field strength $\delta A/B$ at weak (left) and strong (right) SOC strength $\alpha$ in the (a) topological and (b) trivial phases. The spectral functions are plotted along the lower superconducting band $\omega=-E_{-}(\kb)$ at each momentum $\kb$. 
The exchange coupling is set to $h=2$ and $h=0.5$ in (a) and (b), whereas the SOC strength is set to $\alpha=1$ and $\alpha=4$ in the left and right panels, respectively. 
Note that the normalized BC contribution to the spectral function $\frac{\delta A/B}{A_0}$ is directly given by the BC contribution to phase-space DOS per field strength $\delta D/B$, as shown in the insets.}  
}
\label{fig:4}
\end{figure}

Moreover, this topological signature 
can be amplified quadratically by the Rashba SOC strength $\alpha$ since Eq. \ref{eq:deltaDk0} indicates that $\delta A/A_0=\delta D\propto \alpha^2$ at $k=0$.  
We demonstrate this drastic change in $\alpha$ at small momenta in Fig.~\ref{fig:4} (a) and (b) 
in the topological and trivial phases, respectively, 
along with the bare spectral function $A_0(\kb,\w)$ for the lower superconducting band $\omega=-E_{-}(\kb)$. 
In contrast to the small momenta regime, the large momenta regime does not show clear monotonic dependence in $\alpha$.     
Besides amplifying the topological signature, this $\alpha^2$-enhancement at $k=0$ can be used to quantitatively determine the relative SOC strength between two different Rashba layers or interfaces within otherwise identical heterostructures.  

\begin{figure}[t!]
\centering
\includegraphics[width=1\linewidth]{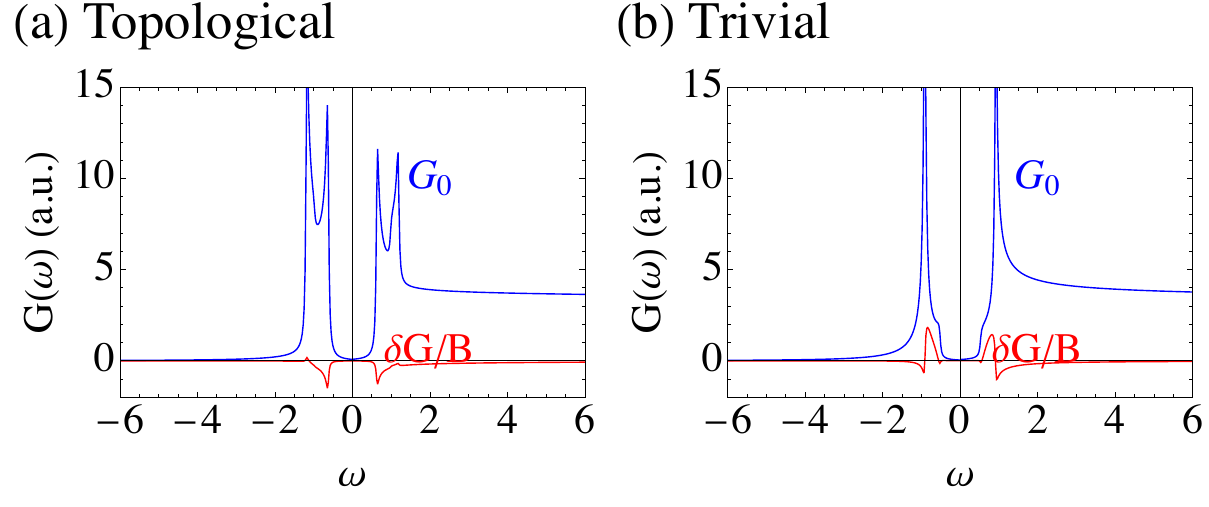}
\caption{
The BC effects in the tunneling conductance $G(\omega)$ in the (a) topological phase $h=2$ and (b) the trivial case with $h=0.5$. The blue curves $G_0$ are the bare tunneling conductance obtained with $D(\textbf{k})=1$, i.e. at $B=0$. The red curves $\delta G/B$ are the corrections per unit magnetic field $B$ to $G_0$ from the BC-induced change in the phase-space DOS, calculated from Eq. \ref{eq:NI} with $D(\textbf{k})$ replaced by $\delta D(\kb)$. We set the SOC strength to be $\alpha=1$ as in Fig \ref{fig:2}. }	
\label{fig:1}
\end{figure}

In contrast to the spectral function, the BC-modified tunneling conductance $G(\omega)=G_0(\omega)+\delta G(\omega)$ is not an effective topological diagnostic. We contrast the results in the topological and trivial phases in Fig.~\ref{fig:1}, where  we omit the position dependence since the signal is nearly uniform at the scale of the wavepacket. Here, the bare conductance $G_0$ and the BC correction $\delta G$ are plotted using Eq. \ref{eq:N} by taking $D(\kb)$ to be $1$ and $\delta D(\kb)$, respectively. 
The bare conductance $G_0$ shows four coherence peaks at frequencies $\omega$ marked by dotted lines, where the peak intensities are not particle-hole symmetric since the effective pairing symmetry is not $s$-wave\cite{Schrieffer}. 
These peaks originate from the band extrema of the lower BdG band (see Fig. \ref{fig:0}(c)), which are accessible in the low-density regime. 
In experiments, the number of visible coherence peaks is determined by the normal state band structure,  thermal and impurity-caused smearing, and the focused frequency range. The bare signal $G_0$ with no BC effect is given by the tunneling spectrum at zero field $\vex{B}=0$. 

When a weak magnetic field \vex{B} is applied, the BC contribution $\delta G$ considerably suppresses the strengths of the two lower-energy coherence peaks but \textit{not} the higher-energy peaks in Fig.~\ref{fig:1}(a). 
As we tune the strengths of the exchange coupling $h$ and SOC $\alpha$ within the topological regime, we generally find a peak suppression or/and enhancement in intensities that are \textit{non-uniform in frequency} $\omega$ and grows linearly with an applied field $B$ weaker than the critical field. 
{Nonetheless, the information about the topological conditions in Eq. \ref{eq:deltaDk0} becomes convoluted in $G(\omega)$ due to the momentum integral in Eq.\ref{eq:N}. 
Therefore, the distinction between the BC effects in the topological and trivial phases seems quantitative or qualitative in a complicated way (see Fig.~\ref{fig:1}(b)). 
Despite that the distinction is quantitative in $G(\omega)$ between topological and trivial phases, our analyses point out that one should expect such B-field-activated nonuniform modulation in experiments due to the phase-space BC. Such BC effect is most easily detectable in lightly doped TSC candidate materials where multiple coherence peaks are visible. }

\section{Acknowledgment}

Y.L. and Y.-T.H. are grateful for the very helpful discussions with Jay Sau. Y.-T.H. also acknowledges helpful discussion with Jhih-Shih You at the initial stage of the project. 
Y.L. acknowledges a postdoctoral
fellowship from the Simons Foundation “Ultra-Quantum Matter” Research Collaboration.
Y.-T.H. acknowledges support from Department of Energy Basic Energy Science Award DE-SC0024291. Y.-T.H. acknowledges support from NSF Grant No. DMR-2238748. 
This work was performed in part at Aspen Center for Physics, which is supported by National Science Foundation grant PHY-2210452. This work was supported in part by the National Science Foundation under Grant No. NSF PHY-1748958. 

\appendix

\section{Wave Packet Properties}\label{app:wave}

In this appendix, we derive some of the key properties of quasiparticle wavepacket discussed in the main text.
The wavepacket center coordinate in the real space defined in Eq.~\ref{eq:rc}~\cite{LiangPRB2017} can be expressed in terms of the coherence factors as 
   \begin{align}\label{Seq:rc0}
 \begin{aligned}
 \rb_c
 =\,&
 \partial_{\kb_c} \theta_{w}(\kb_c, t)
 \\
 &
 +
 i\sum_{\sigma}
 \left( 
 u_{\sigma s}^*(\kb_c) \partial_{\kb_c} u_{\sigma s}(\kb_c)
 +
 v_{\sigma s}^*(\kb_c) \partial_{\kb_c} v_{\sigma s}(\kb_c)
 \right).
\end{aligned}
\end{align}
Here, $\theta_{w}$ stands for the phase of the envelope function  $\theta_{w}(\kb, t) \equiv -\arg w_s(\kb,t)$.
We have used the fact that the integral of any smooth function $f(\kb)$ weighted by $|w_s(\kb,t)|^2 $, which is sharply peaked around the center $\kb_c$, can be approximated  by the value of $f(\kb)$ at  $\kb=\kb_c$:
	\begin{align}
	\int \frac{d^2\kb}{(2\pi)^2} |w_s(\kb,t)|^2 f(\kb)\approx f(\kb_c).
	\end{align} 
Using Eq.~\ref{eq:Aks-1}, Eq.~\ref{Seq:rc0} can be rewritten as
\begin{align}\label{Seq:rc}
\begin{aligned}
\rb_c
=\,&
 \partial_{\kb_c} \theta_{w}(\kb_c, t)
+
\mathcal{A}_{ks}
\\
 =\,&
\partial_{\kb_c} \theta_{w}(\kb_c, t)
-\frac{1}{2}\rho_{0s} (\kb_c) \partial_{\kb_c}  \chi
+\frac{1}{2}\rho_{1s} (\kb_c) \partial_{\kb_c}  \phi_{\kb_c}.
\end{aligned}
\end{align}


Let us now consider the wavepacket average of the total charge $Q$, total spin $S$, and the charge dipole moment $\vex{d}$:
\begin{align}
\begin{aligned}
&
O
=\,\bra{W_s} 
\hat{O}
\ket{W_s} 
-
\bra{G} 
\hat{O}
\ket{G}, 
\qquad
O=Q, S, \vex{d},
\end{aligned}
\end{align}
where $\hat{O}$ represents
\begin{align}
\begin{aligned}
&
\hat{O}=\,
\sum_{\sigma}
\int d \rb 
g_O(\rb)
c_{\sigma }^{\dagger}(\rb)
c_{\sigma }(\rb),
\\
&
g_O(\rb)
=
\begin{cases}
1,
& 
O=Q,
\\
\zeta_{\sigma}/2,
&
O=S,
\\
\rb-\rb_c,
&
O=\vex{d}.
\end{cases}
\end{aligned}
\end{align}
In terms of the quasiparticle operator $\gamma_{s}(\kb)$,  $\hat{O}=\hat{Q},\hat{S}, \hat{\vex{d}}$  can be rewritten as
\begin{widetext}
	
\begin{align}
\begin{aligned}
&\hat{O}
=\,
\sum_{\sigma,s_1,s_2}
\int_{\rb,\kb_1,\kb_2}
\!\!\!\!\!
g_O(\rb)
e^{-i(\kb_1-\kb_2)\cdot \rb}
\left[ 
u_{\sigma s_1}^*(\kb_1)
u_{\sigma s_2} (\kb_2)
\gamma_{s_1}^{\dagger} (\kb_1)
\gamma_{s_2} (\kb_2)
+
v_{\sigma s_1} (-\kb_1)
v_{\sigma s_2}^* (-\kb_2)
\gamma_{s_1} (-\kb_1)
\gamma_{s_2}^{\dagger}(-\kb_2)
\right] 
+...,
\end{aligned}
\end{align}
where $...$ represents terms involving $\gamma\gamma$ or $\gamma^{\dagger}\gamma^{\dagger}$  whose contribution vanishes after taking the expectation value.

With the help of the following identities

\begin{align}\label{eq:gg}
\begin{aligned}
\bra{W_s}
\gamma_{ s_1}^{\dagger}(\kb_1)
\gamma_{s_2} (\kb_2)
\ket{W_s}
=\,&
\int_{\kb_1',\kb_2'} 
w_s^*(\kb_1',t)w_s(\kb_2',t)
\bra{G} 
\gamma_{s}(\kb_1')
\gamma_{ s_1}^{\dagger}(\kb_1)
\gamma_{s_2} (\kb_2)
\gamma_{s}^{\dagger}(\kb_2')
\ket{G}
\\
=\,&
w_s^*(\kb_1,t)w_s(\kb_2,t)
\delta_{s,s_2}
\delta_{s,s_1},
\\
\bra{W_s}
\gamma_{ s_1}(\kb_1)
\gamma_{s_2}^{\dagger} (\kb_2)
\ket{W_s}
=\,&
\int_{\kb_1',\kb_2'} 
w_s^*(\kb_1',t)w_s(\kb_2',t)
\bra{G} 
\gamma_{s}(\kb_1')
\gamma_{ s_1}(\kb_1)
\gamma_{s_2}^{\dagger} (\kb_2)
\gamma_{s}^{\dagger}(\kb_2')
\ket{G}
\\
=\,&
\delta_{\kb_1,\kb_2}
\delta_{s_1,s_2}
-
\delta_{s,s_1}
\delta_{s,s_2}
w_s^*(\kb_2,t)w_s(\kb_1,t).
\end{aligned}
\end{align}
we find
\begin{align}\label{eq:ave}
\begin{aligned}
\bra{W_s} 
\hat{O}
\ket{W_s} 
-
\bra{G} 
\hat{O}
\ket{G} 
=\,&
\sum_{\sigma}
\int_{\rb}
\int_{\kb_1,\kb_2}
e^{-i(\kb_1-\kb_2)\cdot \rb}
g_O(\rb)
\\
&\times
\left[ 
w_s^*(\kb_1,t)
w_s(\kb_2,t)
u_{\sigma s}^*(\kb_1)
u_{\sigma s} (\kb_2)
-
w_s^*(-\kb_2,t)
w_s(-\kb_1,t)
v_{\sigma s} (-\kb_1)
v_{\sigma s}^* (-\kb_2)
\right] .
\end{aligned}
\end{align}
For $\hat{O}=\hat{Q}$ or $\hat{S}$, it is easy to see that the equation above can be further simplified to 
\begin{align}
\begin{aligned}
&
Q
=\,
\int_{\kb}
|w_s(\kb,t)|^2
\rho_{0s}(\kb)
=
\rho_{0s}(\kb_c)
,
\\
&
S
=\,
\frac{1}{2}
\int_{\kb}
|w_s(\kb,t)|^2
\rho_{1s}(\kb)
=
\frac{1}{2}
\rho_{1s}(\kb_c).
\end{aligned}
\end{align}
For the charge dipole moment $\vex{d}$, one can express $\rb e^{-i(\kb_1-\kb_2)\cdot \rb}$  in the Eq.~\ref{eq:ave} as $-i\partial_{\kb_2}e^{-i(\kb_1-\kb_2)\cdot \rb}$, which results in
\begin{align}\label{Seq:d-1}
\begin{aligned}
&\bra{W_s} 
\hat{\vex{d}}
\ket{W_s} 
-
\bra{G} 
\hat{\vex{d}}
\ket{G} 
=\,
\sum_{\sigma}
\int_{\rb}
\int_{\kb_1,\kb_2}
\left( -i\partial_{\kb_2}
e^{-i(\kb_1-\kb_2)\cdot \rb}
\right) 
w_s^*(\kb_1,t)
w_s(\kb_2,t)
\left[ 
u_{\sigma s}^*(\kb_1)
u_{\sigma s} (\kb_2)
-
v_{\sigma s}^* (\kb_1)
v_{\sigma s} (\kb_2)
\right] 
-\rb_c Q
\\
=\,&
i\sum_{\sigma}
\int_{\kb}
w_s^*(\kb,t)
(
\partial_{\kb}
w_s(\kb,t)
)
\left( 
|u_{\sigma s}(\kb )|^2
-
|v_{\sigma s}(\kb )|^2
\right) 
+i
\sum_{\sigma}
\int_{\kb}
|w_s(\kb,t)|^2
\left( 
u_{\sigma s}^*(\kb )
\partial_{\kb}
u_{\sigma s}(\kb )
-
v_{\sigma s}^*(\kb)
\partial_{\kb}
v_{\sigma s}(\kb)
\right) 
-\rb_c Q
\\
=\,&
\int_{\kb}
|w_s(\kb,t)|^2
\left( 
\rho_{0s}(\kb)
\partial_{\kb}
\theta_{w}(\kb,t)
-
\partial_{\kb} \chi/2
+
\rho_{2 s}(\kb)
\partial_{\kb} \phi_{\kb}/2
\right) 
-\rb_c Q
\\
&
+
\frac{i}{2}
\sum_{\sigma}
\int_{\kb}
(
\partial_{\kb}
|w_s(\kb,t)|^2
)
\left( 
|u_{\sigma s}(\kb )|^2
-
|v_{\sigma s}(\kb )|^2
\right) 
+\frac{i}{2}
\sum_{\sigma}
\int_{\kb}
|w_s(\kb,t)|^2
\partial_{\kb}
\left( 
|u_{\sigma s}(\kb )|^2
-
|v_{\sigma s}(\kb)|^2
\right) 
\\
=\,&
\rho_{0s}(\kb_c)
\partial_{\kb_c}
\theta_{w}(\kb_c,t)
-
\partial_{\kb_c} \chi/2
+
\rho_{2 s}(\kb_c)
\partial_{\kb_c} \phi_{\kb_c}/2
-
\rb_c \rho_{0s}(\kb_c),
\end{aligned}
\end{align}
Inserting the expression for $\rb_c$ in Eq.~\ref{Seq:rc}, we find the wave packet average of the charge dipole moment is given by
\begin{align}\label{Seq:d}
\begin{aligned}
	\vex{d}
	=\,
	\frac{1}{2}
	\left( \rho_{0s}^2(\kb_c)-1\right) 
	\partial_{\kb_c} \chi
	+
	\frac{1}{2}
	\left( \rho_{2 s}(\kb_c)-\rho_{0s}(\kb_c)\rho_{1s} (\kb_c)\right) 
	\partial_{\kb_c} \phi_{\kb_c}.
\end{aligned}
\end{align}
where the first and second terms correspond to $\vex{d}_1$ in Eq.\ref{eq:d1} and $\vex{d}_2$ in Eq.\ref{eq:P-0} respectively.

\section{Lagrangian}\label{app:Lagrangian}
This appendix evaluates the Lagrangian
which governs the semiclassical dynamics of the  wave packet:
 \begin{align}\label{eq:L-0}
\begin{aligned}
	L
	=\,&
	\bra{W_s}
	\left( i \frac{d}{dt} - H\right) 
	\ket{W_s}
	-
	\bra{G}
	\left( i \frac{d}{dt} - H\right) 
	\ket{G}.
\end{aligned}
\end{align}
The part that contains the Hamiltonian is given by
 \begin{align}
\begin{aligned}
	\bra{W_s}
	H
	\ket{W_s}
	-
	\bra{G}
	H
	\ket{G}
	=\,&
	\int_{\kb_1,\kb_2} 
	w_s^*(\kb_1,t)w_s(\kb_2,t)
	\bra{G}
	\gamma_{s}(\kb_1)
	\left( 
	\int_{\kb}\sum_{s'} E_{s'} (k)
	\gamma_{s'}^{\dagger}(\kb)
	\gamma_{s'}(\kb)
	\right) 
	\gamma_{s}^{\dagger}(\kb_2)
	\ket{G}
	\\
	=\,&
	\int_{\kb} 
	|w_s(\kb,t)|^2
	E_{s}(k) 
	=
	E_{ s}(k_c),
\end{aligned}
\end{align}
while the part that involves the time derivative reduces to
 \begin{align}\label{eq:L-1}
\begin{aligned}
\bra{W_s}
i \frac{d}{dt}
\ket{W_s}
=\,&
i
\int_{\kb} 
w_s^*(\kb,t)  \frac{\partial}{\partial t}w_s(\kb,t)
+
i\frac{d \rb_c}{dt} \cdot
\int_{\kb} 
w_s^*(\kb,t)  
\frac{\partial}{\partial \rb_c}
w_s(\kb,t)
\\
&+\,
i \frac{d \rb_c}{dt} \cdot
\int_{\kb_1,\kb_2} 
w_s^*(\kb_1,t)w_s(\kb_2,t)
\bra{G}
\gamma_{s}(\kb_1)
(
\frac{\partial}{\partial \rb_c} 
\gamma_{s}^{\dagger}(\kb_2)
)
\ket{G}.
\end{aligned}
\end{align}
Note that the time dependence of the quasiparticle creation operator $\gamma^{\dagger}_s$
comes from the time dependence of the wave-packet center $\rb_c$. 

Using the BdG transformation Eq.~\ref{eq:BdG}, the term $\bra{W_s}
i \frac{d}{dt}
\ket{W_s}$ can be further divided into three parts:
\begin{align}
\begin{aligned}
&
\bra{W_s}
i \frac{d}{dt}
\ket{W_s}
=\,L_1+L_2+L_3,
\\
&L_1
=\,
i
\int_{\kb} 
w_s^*(\kb,t)  \frac{\partial}{\partial t}w_s(\kb,t)
+
i
\frac{d \rb_c}{dt} \cdot
\int_{\kb} 
w_s^*(\kb,t) 
\frac{\partial}{\partial \rb_c}
w_s(\kb,t)
\\
&
L_2
=\,
i \frac{d \rb_c}{dt} \cdot
\int_{\kb_1,\kb_2} 
w^*_{s}(\kb_1,t)w_s(\kb_2,t)
\\
&\times
\bra{G}
\gamma_{s}(\kb_1)
\left[
(\frac{\partial}{\partial \rb_c}
u_{\uparrow s}(\kb_2)
)
c_{\uparrow}^{\dagger}(\kb_2)
+
(\frac{\partial}{\partial \rb_c}
u_{\downarrow s}(\kb_2)
)
c_{\downarrow}^{\dagger}(\kb_2)
+
(\frac{\partial}{\partial \rb_c}
v_{\downarrow s}(\kb_2)
)
c_{\downarrow}(-\kb_2)
-
(\frac{\partial}{\partial \rb_c}
v_{\uparrow s }(\kb_2)
)
c_{\uparrow}(-\kb_2)
\right]
\ket{G},
\\
&L_3
=\,
i \frac{d \rb_c}{dt} \cdot
\int_{\kb_1,\kb_2} 
w^*_{s}(\kb_1,t)w_s(\kb_2,t)
\\
&\times
\bra{G}
\gamma_{s}(\kb_1)
\left[ 
u_{\uparrow s} (\kb_2)
(\frac{\partial}{\partial \rb_c}
c_{\uparrow}^{\dagger}(\kb_2)
)
+
u_{\downarrow s}(\kb_2)
(\frac{\partial}{\partial \rb_c}
c_{\downarrow}^{\dagger}(\kb_2)
)
+
v_{\downarrow s}(\kb_2)
(\frac{\partial}{\partial \rb_c}
c_{\downarrow}(-\kb_2)
)
-
v_{\uparrow s}(\kb_2)
(\frac{\partial}{\partial \rb_c}
c_{\uparrow}(-\kb_2)
)
\right] 
\ket{G}.
\end{aligned}
\end{align}

Here $L_1$ can be expressed in terms of the phase of the envelope function $w_s(\kb,t)$ at the wave packet center $\kb_c$, i.e., $\theta_{w}(\kb_c,t)$,
\begin{align}
\begin{aligned}
L_1
=\,&
\int_{\kb} 
|w_s(\kb,t)|^2  
(
\frac{\partial}{\partial t} \theta_{w}(\kb,t)
+
\frac{d\rb_c}{d t}
\cdot
\frac{\partial}{\partial \rb_c} \theta_{w}(\kb,t))
\\
=\,&
 \frac{\partial}{\partial t} \theta_{w}(\kb_c,t)
 +
\frac{d\rb_c}{d t}
\cdot
\frac{\partial}{\partial \rb_c } \theta_{w}(\kb_c,t),
\end{aligned}
\end{align}
Using the expression for the wave-packet center $\rb_c$ in Eq.~\ref{Seq:rc}, one can rewrite this equation as
\begin{align}
\begin{aligned}
	L_1=\,&
	\frac{d}{d t} \theta_{w}(\kb_c,t)
	-
	\frac{d\kb_c}{d t}
	\cdot
	\frac{\partial}{\partial \kb_c} \theta_{w}(\kb_c,t)
    \\
	=\,&
	\frac{d}{d t} \theta_{w}(\kb_c,t)
	-\dot{\kb}_c
	\cdot
	\left( 
	\rb_c
	-
	\mathcal{A}_{ks}
	\right).
\end{aligned}
\end{align}

Converting the electron operator $c_{\sigma}$  into the quasiparticle operator $\gamma_{s}$ in $L_2$, we obtain
 \begin{align}\label{eq:L2}
\begin{aligned}
&L_2
=\,
i \frac{d \rb_c}{dt} \cdot
\int_{\kb_1,\kb_2} 
w_s^*(\kb_1,t)w_s(\kb_2,t)
\bra{G}
\gamma_{s}(\kb_1)
\left[
\begin{aligned}
&
\left( 
\frac{\partial}{\partial \rb_c}
u_{\uparrow s}(\kb_2)
\right) 
\sum_{s'}
\left( 
u_{\uparrow s'}^*(\kb_2)
\gamma_{s'}^{\dagger}(\kb_2)
-
v_{\uparrow s'}(-\kb_2)
\gamma_{s'}(-\kb_2)
\right) 
\\
&
+
\left( \frac{\partial}{\partial \rb_c}
u_{\downarrow s} (\kb_2)
\right) 
\sum_{s'}
\left( 
u_{\downarrow s'}^*(\kb_2)
\gamma_{s'}^{\dagger}(\kb_2)
+
v_{\downarrow s'}(-\kb_2)
\gamma_{s'}(-\kb_2)
\right) 
\\
&
+
\left( \frac{\partial}{\partial \rb_c}
v_{\downarrow s}(\kb_2)
\right) 
\sum_{s'}
\left( 
v_{ \downarrow s'}^*(\kb_2)
\gamma_{s'}^{\dagger}(\kb_2)
+
u_{\downarrow s'}(-\kb_2)
\gamma_{ s'}(-\kb_2)
\right) 
\\
&
+
\left( 
\frac{\partial}{\partial \rb_c}
v_{\uparrow s}(\kb_2)
\right) 
\sum_{s'}
\left( 
v_{\uparrow s'}^*(\kb_2)
\gamma_{s'}^{\dagger}(\kb_2)
-
u_{\uparrow s'}(-\kb_2)
\gamma_{s'}(-\kb_2)
\right) 
\end{aligned}
\right]
\ket{G}
\\
=\,&
i \frac{d \rb_c}{dt} \cdot
\int_{\kb} 
|w_s(\kb,t)|^2
\sum_{\sigma}
\left( 
u_{\sigma s}^*(\kb)
\frac{\partial}{\partial \rb_c}
u_{\sigma s} (\kb)
+
v_{\sigma s}^*(\kb)
\frac{\partial}{\partial \rb_c}
v_{\sigma s}(\kb)
\right)
\\
=\,&
\dot{\rb}_c \cdot
\mathcal{A}_{rs}. 
\end{aligned}
\end{align}


For the evaluation of $L_3$, note that
\begin{align}
\begin{aligned}
\frac{d \rb_c}{dt} \cdot
\frac{\partial}{\partial \rb_c}
&c_{\sigma}(\pm \kb)
=\,
\frac{d \rb_c}{dt} \cdot
\frac{\partial}{\partial \rb_c}
\left( 
\int d^2 \rb
e^{-i(\pm \kb+\vex{A}(\rb_c))\cdot \rb}
c_{\sigma}(\rb)
\right) 
\\
=\,&
-i \frac{d \rb_c}{dt} \cdot
\int_{\rb}
\frac{\partial \vex{A}(\rb_c) \cdot \rb }{\partial \rb_c}
e^{-i(\pm \kb+\vex{A}(\rb_c))\cdot \rb}
c_{\sigma}(\rb)
\\
=\,&
\frac{d \vex{A}(\rb_c)}{dt} \cdot
\int d^2 \rb
\left( 
\pm 
\frac{\partial }{\partial \kb}
e^{-i(\pm \kb+\vex{A}(\rb_c))\cdot \rb}
\right) 
c_{\sigma}(\rb)
\\
=\,&
\pm 
\frac{d \vex{A}(\rb_c)}{dt} \cdot
\frac{\partial }{\partial \kb}
c_{\sigma} (\pm \kb),
\end{aligned}
\end{align}
which leads to

\begin{align}
\begin{aligned}
L_3=\,&
i \frac{d}{dt} \vex{A}(\rb_c) \cdot
\int_{\kb_1,\kb_2} 
w_s^*(\kb_1,t)w_s(\kb_2,t)
\\
&\times
\bra{\Omega}
\gamma_{s} (\kb_1)
\left[ 
u_{\uparrow s}(\kb_2)
(\frac{\partial }{\partial \kb_2}
c_{\uparrow}^{\dagger}(\kb_2)
)
+
u_{\downarrow s}(\kb_2)
(\frac{\partial }{\partial \kb_2}
c_{\downarrow}^{\dagger}(\kb_2)
)
-
v_{\downarrow s}(\kb_2)
(\frac{\partial }{\partial \kb_2}
c_{\downarrow}(-\kb_2)
)
+
v_{\uparrow s}(\kb_2)
(\frac{\partial }{\partial \kb_2}
c_{\uparrow}(-\kb_2)
)
\right] 
\ket{\Omega}
\\
=\,&
-i \frac{d}{dt}\vex{A}(\rb_c) \cdot
\int_{\kb} 
w_s^*(\kb,t) \left(  \partial_{\kb} w_s(\kb,t) \right) 
\left( 
|u_{\sigma s}(\kb)|^2
-
|v_{\sigma s}(\kb)|^2
\right) 
\\
&
-i \frac{d }{dt} \vex{A}(\rb_c) \cdot
\int_{\kb} 
|w_s(\kb,t)|^2
\left( 
u_{\sigma s}^*(\kb)
\partial_{\kb}
u_{\sigma s} (\kb)
-
v_{ \sigma s}^*(\kb)
\partial_{\kb}
v_{\sigma s}(\kb)
\right) .
\end{aligned}
\end{align}

Comparing this equation with the expression for the electric dipole moment $\vex{d}$ (See Eqs.~\ref{Seq:d-1} and~\ref{Seq:d}), it is easy to see that $L_3$ can be expressed as
\begin{align}
\begin{aligned}
L_3
=\,-\dot{\vex{A}} (\rb_c) \cdot (\rho_{0s}(\kb_c) \rb_c+\vex{d}).
\end{aligned}
\end{align}

Combining everything, we obtain
\begin{align}
\begin{aligned}
L
=\,&
-
E_{s}(k_c)
-
\dot{\kb}_c
\cdot
\left( 
\rb_c
-
\mathcal{A}_{ks}
\right)
+
\dot{\rb}_c \cdot
\mathcal{A}_{rs} 
\\
&
-
\dot{\vex{A}}(\rb_c) \cdot (\rho_{0s}(\kb_c) \rb_c+\vex{d}).
\end{aligned}
\end{align}
Here we have dropped terms which can be expressed as total time derivatives and as a result do not enter the equation of motion.
Using $\vex{A} (\rb_c)=\frac{1}{2} \vex{B} \times \rb_c$, we find that the Lagrangian can be rewritten as Eq.\ref{eq:L}.

\section{Equation of Motion}\label{app:EOM}

In this appendix, we derive the explicit expression for the equation of motion of the quasiparticle wavepacket dynamics in terms of the coherence factors. 
Inserting the expression for the Lagrangian in Eq.~\ref{eq:L} into the Euler–Lagrange equation
\begin{align}
\begin{aligned}
	\frac{\partial L}{\partial r^{i}_c}
	=
	\frac{d}{dt} \frac{\partial L}{\partial \dot{r}^{i}_c},
	\qquad
	\frac{\partial L}{\partial k^{i}_c}
	=
	\frac{d}{dt} \frac{\partial L}{\partial \dot{k}^{i}_c},
	\qquad
	i=x,y,
\end{aligned}
\end{align}
one can derive the equation of motion for the wave-packet center:
\begin{align}\label{eq:EOM-1}
\begin{aligned}
\dot{k}^{i}_c
=\,&
-\partial_{r^{i}_c} E_{s} (k_c)
+
\Omega_{rks}^{ij}
\dot{k}^{j}_c
+
\Omega_{rrs}^{ij}
\dot{r}^{j}_c ,
\\
\dot{r}^{i}_c
=\,&
\partial_{k^{i}_c} E_{ s} (k_c)
-
\Omega_{kks}^{ij}
\dot{k}^{j}_c
-
\Omega_{krs}^{ij}
\dot{r}^{j}_c.
\end{aligned}
\end{align}
Here  we have employed the notation that repeated indices implies summation.
Note that the Berry curvatures $\Omega$ are defined as derivatives of the Berry connections $\mathcal{A}_{ks}$ and ${\mathcal{A}}_{rs}$:
\begin{align}\label{eq:Omega-0}
\begin{aligned}
	\Omega_{kks}^{ij}
	\equiv\,&
	 \partial_{k^{i}_c} {\mathcal{A}}_{ks}^{j}
	-\partial_{k^{j}_c} {\mathcal{A}}_{ks}^{i},
	\qquad
	\Omega_{rrs}^{ij}
	\equiv\,
	\partial_{r^{i}_c} {{\mathcal{A}}}_{rs}^{j}
	-\partial_{r^{j}_c} {{\mathcal{A}}}_{rs}^{i},
		\\
	\Omega_{rks}^{ij}
	\equiv \,&
	-\Omega_{krs}^{ji}
	\equiv \,
	\partial_{r^{i}_c} {\mathcal{A}}_{ks}^{j}
	-\partial_{k^{j}_c} {{\mathcal{A}}}_{rs}^{i}.
\end{aligned}
\end{align}

Eq.~\ref{eq:EOM-1} can be rewritten in a vector form:
\begin{align}\label{eq:EOM-2}
\begin{aligned}
\dot{\kb}_c
=\,&
-
\nabla_{\rb_c}  E_{s}(k_c)
+
\dot{\rb}_c
\times
{\bf \Omega}_{rs}
+
\nabla_{\rb_c}
\left( 
\dot{\kb}_c
\cdot
{\bf \mathcal{A}}_{ks}
\right) 
-
\left( 
\dot{\kb}_c
\cdot
\nabla_{\kb_c}
\right) 
{\bf {\mathcal{A}}}_{rs},
\\
\dot{\rb}_c
=\,&
\nabla_{\kb_c}  E_{s} (k_c)
-
\dot{\kb}_c
\times
{\bf \Omega}_{ks}
-
\nabla_{\kb_c}
\left( 
\dot{\rb}_c
\cdot
{\bf {\mathcal{A}}}_{rs}
\right) 
+
\left( 
\dot{\rb}_c
\cdot
\nabla_{\rb_c}
\right) 
{\bf \mathcal{A}}_{ks}
.
\end{aligned}
\end{align}
Here the vector forms of the Berry curvatures ${\bf \Omega}_{rs}$ and ${\bf \Omega}_{ks}$ are defined as
\begin{align}
\begin{aligned}
&{\bf \Omega}_{ks}
=\,
\nabla_{\kb_c} \times \mathcal{A}_{ks},
\qquad
{\bf \Omega}_{rs}
=\,
\nabla_{\rb_c} \times {\mathcal{A}}_{rs},
\end{aligned}
\end{align}
and they are related to their tensor counterparts $\Omega_{rrs}$ and $\Omega_{kks}$ by:
\begin{align}
\begin{aligned}
{\bf \Omega}^{l}_{rs/ks}= & \frac{1}{2}\epsilon_{lij} \Omega^{ij}_{rrs/kks},
\qquad
\Omega^{ij}_{rrs/kks}= & \epsilon_{ijl} {\bf \Omega}^{l}_{rs/ks},
\end{aligned}
\end{align}
with $\epsilon_{ijl}$ being the Levi-Civita symbol.
Note that in the present paper, we consider a 2D system, and therefore only the $z$ component of the vector Berry curvature ${\bf \Omega}$ is nonvanishing. 

Using the expressions for the Berry connections $\mathcal{A}_{ks}$ and ${\mathcal{A}}_{rs}$, we obtain
\begin{align}\label{Seq:vectorOmega}
\begin{aligned}
&{\bf \Omega}_{ks}
=\,
-\frac{1}{2}\nabla_{\kb_c}\rho_{0s} (\kb_c)  \times \nabla_{\kb_c}  \chi
+\frac{1}{2}\nabla_{\kb_c} \rho_{1s} (\kb_c) \times \nabla_{\kb_c}  \phi_{\kb_c},
\\
&{\bf \Omega}_{rs}
=
-
\nabla_{\rb_c}\rho_{0s} (\kb_c) \times 
\left( \nabla_{\rb_c}  \chi/2-\vex{A}(\rb_c) \right) 
+
\rho_{0 s}(\kb_c)\vex{B}
+
\frac{1}{2}
\nabla_{\rb_c} \times (\vex{B} \times \vex{d})
+
\frac{1}{2}\nabla_{\rb_c}\rho_{1s} (\kb_c) \times \nabla_{\rb_c}  \phi_{\kb_c}.
\end{aligned}
\end{align}
The tensor-form of the phase-space Berry curvature $\Omega_{rk}$ is given by,
\begin{align}
\begin{aligned}
	\Omega_{rks}^{ij}
	= \,&
    -\frac{1}{2}
    \left(
    \partial_{r^{i}_c} \rho_{0s}(\kb_c)  \partial_{k^{j}_c}  \chi
    -
    \partial_{k^{j}_c} \rho_{0s}(\kb_c)  \partial_{r^{i}_c}  \chi
    \right)
	+
    \frac{1}{2}
    \left(
    \partial_{r^{i}_c}\rho_{1s}(\kb_c)  \partial_{k^{j}_c}  \phi_{\kb_c}
	-
    \partial_{k^{j}_c}\rho_{1s}(\kb_c)  \partial_{r^{i}_c}  \phi_{\kb_c}
    \right)
    \\
    &
    -
    \partial_{k^{j}_c} \rho_{0s}(\kb_c)  \vex{A}^{i}(\rb_c)
    -
    \frac{1}{2}\partial_{k^{j}_c} (\vex{B}\times \vex{d})^{i}.
\end{aligned}
\end{align}

Inserting Eq.~\ref{Seq:vectorOmega} into  Eq.~\ref{eq:EOM-2} leads to the following form of the equation of motion

\begin{align}\label{eq:EOM-3}
\begin{aligned}
\dot{\kb}_c
=\,&
-
\nabla_{\rb_c}  E_{ s} (k_c)
+
\dot{\rb}_c
\times
\left[ 
-
\nabla_{\rb_c}\rho_{0s} (\kb_c) \times \left( \frac{1}{2}\nabla_{\rb_c}  \chi-  \vex{A}(\rb_c) \right) 
+
\rho_{0 s}(\kb_c) \vex{B}
+
\frac{1}{2}
\nabla_{\rb_c} \times (\vex{B} \times \vex{d})
+
\frac{1}{2}\nabla_{\rb_c}\rho_{1s} (\kb_c) \times \nabla_{\rb_c}  \phi_{\kb_c}
\right] 
\\
&
-
\frac{1}{2}
\nabla_{\rb_c}
\left[ 
\dot{\kb}_c
\cdot
\left( 
\rho_{0s} (\kb_c) \nabla_{\kb_c}  \chi
-\rho_{1s} (\kb_c) \nabla_{\kb_c}  \phi_{\kb_c}
\right) 
\right] 
+
\left( 
\dot{\kb}_c
\cdot
\nabla_{\kb_c}
\right) 
\left[ 
\rho_{0s} (\kb_c) \left(\frac{1}{2} \nabla_{\rb_c}  \chi -\vex{A}(\rb_c)\right) 
-
\frac{1}{2}
\vex{B} \times \vex{d}
-
\frac{1}{2}
\rho_{1s} (\kb_c) \nabla_{\rb_c}  \phi_{\kb_c}
\right] 
,
\\
\dot{\rb}_c
=\,&
\nabla_{\kb_c}  E_{s} (k_c)
-
\dot{\kb}_c
\times
\left( 
-\frac{1}{2}\nabla_{\kb_c}\rho_{0s} (\kb_c)  \times \nabla_{\kb_c}  \chi
+\frac{1}{2}\nabla_{\kb_c} \rho_{1s} (\kb_c) \times \nabla_{\kb_c}  \phi_{\kb_c}
\right) 
\\
&
+
\nabla_{\kb_c}
\left\lbrace 
\dot{\rb}_c
\cdot
\left[ 
\rho_{0s} (\kb_c) \left( \frac{1}{2}\nabla_{\rb_c}  \chi-\vex{A}(\rb_c)\right) 
-
\frac{1}{2}
\vex{B} \times \vex{d}
-
\frac{1}{2}
\rho_{1s} (\kb_c) \nabla_{\rb_c}  \phi_{\kb_c}
\right] 
\right\rbrace 
-
\frac{1}{2}
\left( 
\dot{\rb}_c
\cdot
\nabla_{\rb_c}
\right) 
\left( 
\rho_{0s} (\kb_c) \nabla_{\kb_c}  \chi
-\rho_{1s} (\kb_c) \nabla_{\kb_c}  \phi_{\kb_c}
\right) 
.
\end{aligned}
\end{align}
 where the dependence of $\rho_{as}$ on the coherence factors is given in Eqs.~\ref{eq:rho} and~\ref{eq:rho2}.

\section{Berry Curvature Correction to the Phase-space Density of States}\label{app:DOS}

As mentioned in the main text,   the equation of motion Eq.~\ref{eq:EOM-1}  exhibits a noncanonical structure due to the presence of the Berry curvatures, which leads to the breakdown of  the conservation of the phase-space volume. 
In particular, the phase-space volume element $\Delta V=\Delta\vex{k} \Delta \vex{r}$ is no longer constant in time and instead evolves  according to~\cite{NiuDOS}
\begin{align}
\begin{aligned}
	\frac{1}{\Delta V}
	\dfrac{d \Delta V}{d t}
	=
	\nabla_{\rb} \cdot \dot{\rb} + \nabla_{\kb} \cdot \dot{\kb}.
\end{aligned}
\end{align}
One therefore needs to introduce a modified phase-space density of states $D(\rb,\kb)$ such that the number of states in volume element $\Delta V$, i.e., $D(\rb,\kb)\Delta V$, remains constant over time.
This appendix provides the derivation of the correction to the phase-space density of states induced by the Berry curvatures for the current model.

In terms of the Berry curvatures, up to an inessential coefficient, the modified  phase-space density of states $D(\rb,\kb)$ acquires a form~\cite{NiuRev,Arnold}
\begin{align}\label{eq:D-1}
\begin{aligned}
D
=&
\sqrt{
	\det 
	\begin{bmatrix}
	\Omega_{rrs}  
	&
	\Omega_{rks}-I
	\\
	\Omega_{krs}+I
	&
	\Omega_{kks}
	\end{bmatrix}
}.
\end{aligned}
\end{align}
Since the external perturbation is assumed to be slowly varying in real space, one can perform an expansion in terms of the spatial gradient and keep only the leading order terms~\cite{NiuSC}.
The Berry curvature correction to the phase-space density of states $\delta D \equiv D-1$ therefore can be approximated by
\begin{align}\label{eq:D}
\begin{aligned}
& \delta D(\rb,\kb)
=
\Tr \Omega_{krs}
-
{\bf \Omega}_{rs} \cdot {\bf \Omega}_{ks}
\\
&=
-
\nabla_{\kb} \rho_{0s}  \cdot \vex{p}_s
-
\frac{1}{2}
\vex{B}  \cdot (\nabla_{\kb} \times \vex{d} )
+
\frac{1}{2}
\nabla_{\rb} \rho_{0s}   \cdot \nabla_{\kb}  \chi
+
\frac{1}{2}
\nabla_{\kb}  \rho_{1s}  \cdot \nabla_{\rb}   \phi_{\kb}
-
\frac{1}{2}
\nabla_{\rb} \rho_{1s}  \cdot  \nabla_{\kb} \phi_{\kb}
\\
	&
	-
	\frac{1}{2}\left( 
	\nabla_{\kb}\rho_{0s}  \times \nabla_{\kb}  \chi
	-\nabla_{\kb} \rho_{1s} \times \nabla_{\kb}  \phi_{\kb}
	\right) 
	\cdot
	\left( 
	\nabla_{\rb}\rho_{0s} (\kb) \times 
	\vex{p}_s 
	-
	\rho_{0 s}\vex{B}
	-
	\frac{1}{2}
	\nabla_{\rb} \times (\vex{B} \times \vex{d})
	-
	\frac{1}{2}
	\nabla_{\rb}\rho_{1s} (\kb) \times  \nabla_{\rb} \phi_{\kb}
	\right).
\end{aligned}
\end{align}
where in the second step we have inserted the explicit expressions for the Berry curvatures.
$\vex{p}_s $ denotes the supercurrent 
$
\vex{p}_s 
\equiv 
\frac{1}{2} \nabla_{\rb}   \chi
-
\vex{A}.
$
	
We now separate $\delta D$ into three different components: the external magnetic field $\vex{B}$ dependent part $\delta D_1$,	
the supercurrent $\vex{p}_s$ dependent part $\delta D_2$, 
and the remaining part $\delta D_3$:
\begin{align}
	\begin{aligned}
	&
	\delta D_1(\rb,\kb)
	=
	-
	\frac{1}{2}
	\vex{B}  \cdot (\nabla_{\kb} \times \vex{d} )
	+
	\frac{1}{2}
	\left( \nabla_{\kb}  \rho_{0s} \times  \nabla_{\kb}   \chi
	-\nabla_{\kb} \rho_{1s}  \times \nabla_{\kb}   \phi_{\kb} \right) 
	\cdot
	\left[ 
	\rho_{0s} \vex{B}
	+
	\frac{1}{2}
	\nabla_{\rb} \times \left( \vex{B} \times \vex{d} \right) 
	\right], 
	\\
	&
	\delta D_2(\rb,\kb)
	=
	-
	\nabla_{\kb} \rho_{0s}  \cdot \vex{p}_s
	-
	\frac{1}{2}
	\left( \nabla_{\kb}  \rho_{0s} \times  \nabla_{\kb}   \chi
	-\nabla_{\kb} \rho_{1s}  \times \nabla_{\kb}   \phi_{\kb} \right) 
	\cdot
	\left( 
	\nabla_{\rb}  \rho_{0s} \times \vex{p}_s
	\right) ,
	\\
	&
	\delta D_3(\rb,\kb)
	=
	\frac{1}{2}
	\left( 
	\nabla_{\rb} \rho_{0s}   \cdot \nabla_{\kb}  \chi
	+
	\nabla_{\kb}  \rho_{1s}  \cdot \nabla_{\rb}   \phi_{\kb}
	-
	\nabla_{\rb} \rho_{1s}  \cdot  \nabla_{\kb} \phi_{\kb}
	\right) 
	+
	\frac{1}{4}
	\left( \nabla_{\kb}  \rho_{0s} \times  \nabla_{\kb}   \chi
	-\nabla_{\kb} \rho_{1s}  \times \nabla_{\kb}   \phi_{\kb}\right) 
	\cdot
	\left( 
	\nabla_{\rb} \rho_{1s}  \times \nabla_{\rb}   \phi_{\kb}
	\right) .
	\end{aligned}
	\end{align}

	Dropping the higher order terms in
			the spatial gradient $\nabla_{\rb}$ in the equation above, we find $\delta D_3=0$ and $
            \delta D_2(\rb,\kb)
		=
		-
		\nabla_{\kb} \rho_{0s}  \cdot \vex{p}_s
		$,  which is not considered in this paper. 
        We focus instead on the component $\delta D_1(\rb,\kb)$, which evaluates to
		\begin{align}
        \begin{aligned}
		\delta D_1(\rb,\kb)
		=&
		-
		\frac{1}{2}
		\vex{B}  \cdot (\nabla_{\kb} \times \vex{d} )
		+
		\frac{1}{2}
		\rho_{0s} 
		\left( 
		\nabla_{\kb}  \rho_{0s} \times  \nabla_{\kb}   \chi
		-\nabla_{\kb} \rho_{1s}  \times \nabla_{\kb}   \phi_{\kb}
		\right) 
		\cdot
		\vex{B}
        \\
        	=&
	-
	\frac{1}{4}
	\vex{B}  \cdot
	\left[  
	\left( 
	\nabla_{\kb} \rho_{2s }
	+
	\rho_{0s} 
	\nabla_{\kb} \rho_{1s} 
	-
	\rho_{1s} 
	\nabla_{\kb} \rho_{0s} 
	\right) 
	\times 
	\nabla_{\kb} 
	\phi_{\kb}
	\right].
		\end{aligned}
		\end{align}
	Here in the second equality, we have inserted the expression for $\vex{d}$ in Eq.~\ref{Seq:d}.
    Note that this expression reduces to Eq.~\ref{eq:dD1}
    for s-wave pairing $\chi=0$.
    
      \end{widetext}

\bibliography{BerryCur_2.bib}

\begin{thebibliography}{59}%
\makeatletter
\providecommand \@ifxundefined [1]{%
 \@ifx{#1\undefined}
}%
\providecommand \@ifnum [1]{%
 \ifnum #1\expandafter \@firstoftwo
 \else \expandafter \@secondoftwo
 \fi
}%
\providecommand \@ifx [1]{%
 \ifx #1\expandafter \@firstoftwo
 \else \expandafter \@secondoftwo
 \fi
}%
\providecommand \natexlab [1]{#1}%
\providecommand \enquote  [1]{``#1''}%
\providecommand \bibnamefont  [1]{#1}%
\providecommand \bibfnamefont [1]{#1}%
\providecommand \citenamefont [1]{#1}%
\providecommand \href@noop [0]{\@secondoftwo}%
\providecommand \href [0]{\begingroup \@sanitize@url \@href}%
\providecommand \@href[1]{\@@startlink{#1}\@@href}%
\providecommand \@@href[1]{\endgroup#1\@@endlink}%
\providecommand \@sanitize@url [0]{\catcode `\\12\catcode `\$12\catcode
  `\&12\catcode `\#12\catcode `\^12\catcode `\_12\catcode `\%12\relax}%
\providecommand \@@startlink[1]{}%
\providecommand \@@endlink[0]{}%
\providecommand \url  [0]{\begingroup\@sanitize@url \@url }%
\providecommand \@url [1]{\endgroup\@href {#1}{\urlprefix }}%
\providecommand \urlprefix  [0]{URL }%
\providecommand \Eprint [0]{\href }%
\providecommand \doibase [0]{http://dx.doi.org/}%
\providecommand \selectlanguage [0]{\@gobble}%
\providecommand \bibinfo  [0]{\@secondoftwo}%
\providecommand \bibfield  [0]{\@secondoftwo}%
\providecommand \translation [1]{[#1]}%
\providecommand \BibitemOpen [0]{}%
\providecommand \bibitemStop [0]{}%
\providecommand \bibitemNoStop [0]{.\EOS\space}%
\providecommand \EOS [0]{\spacefactor3000\relax}%
\providecommand \BibitemShut  [1]{\csname bibitem#1\endcsname}%
\let\auto@bib@innerbib\@empty
\bibitem [{\citenamefont {Read}\ and\ \citenamefont {Green}(2000)}]{ReadGreen}%
  \BibitemOpen
  \bibfield  {author} {\bibinfo {author} {\bibfnamefont {N.}~\bibnamefont
  {Read}}\ and\ \bibinfo {author} {\bibfnamefont {D.}~\bibnamefont {Green}},\
  }\bibfield  {title} {\enquote {\bibinfo {title} {Paired states of fermions in
  two dimensions with breaking of parity and time-reversal symmetries and the
  fractional quantum hall effect},}\ }\href {\doibase
  10.1103/PhysRevB.61.10267} {\bibfield  {journal} {\bibinfo  {journal} {Phys.
  Rev. B}\ }\textbf {\bibinfo {volume} {61}},\ \bibinfo {pages} {10267--10297}
  (\bibinfo {year} {2000})}\BibitemShut {NoStop}%
\bibitem [{\citenamefont {Kitaev}(2001)}]{Kitaev2001aa}%
  \BibitemOpen
  \bibfield  {author} {\bibinfo {author} {\bibfnamefont {A.~Y.}\ \bibnamefont
  {Kitaev}},\ }\bibfield  {title} {\enquote {\bibinfo {title} {Unpaired
  majorana fermions in quantum wires},}\ }\href {\doibase
  10.1070/1063-7869/44/10S/S29} {\bibfield  {journal} {\bibinfo  {journal}
  {Physics-Uspekhi}\ }\textbf {\bibinfo {volume} {44}},\ \bibinfo {pages} {131}
  (\bibinfo {year} {2001})}\BibitemShut {NoStop}%
\bibitem [{\citenamefont {Jiao}\ \emph {et~al.}(2020)\citenamefont {Jiao},
  \citenamefont {Howard}, \citenamefont {Ran}, \citenamefont {Wang},
  \citenamefont {Rodriguez}, \citenamefont {Sigrist}, \citenamefont {Wang},
  \citenamefont {Butch},\ and\ \citenamefont {Madhavan}}]{JiaoNature2020}%
  \BibitemOpen
  \bibfield  {author} {\bibinfo {author} {\bibfnamefont {L.}~\bibnamefont
  {Jiao}}, \bibinfo {author} {\bibfnamefont {S.}~\bibnamefont {Howard}},
  \bibinfo {author} {\bibfnamefont {S.}~\bibnamefont {Ran}}, \bibinfo {author}
  {\bibfnamefont {Z.}~\bibnamefont {Wang}}, \bibinfo {author} {\bibfnamefont
  {J.~O.}\ \bibnamefont {Rodriguez}}, \bibinfo {author} {\bibfnamefont
  {M.}~\bibnamefont {Sigrist}}, \bibinfo {author} {\bibfnamefont
  {Z.}~\bibnamefont {Wang}}, \bibinfo {author} {\bibfnamefont {N.~P.}\
  \bibnamefont {Butch}}, \ and\ \bibinfo {author} {\bibfnamefont
  {V.}~\bibnamefont {Madhavan}},\ }\bibfield  {title} {\enquote {\bibinfo
  {title} {Chiral superconductivity in heavy-fermion metal ute2},}\ }\href
  {\doibase 10.1038/s41586-020-2122-2} {\bibfield  {journal} {\bibinfo
  {journal} {Nature}\ }\textbf {\bibinfo {volume} {579}},\ \bibinfo {pages}
  {523--527} (\bibinfo {year} {2020})}\BibitemShut {NoStop}%
\bibitem [{\citenamefont {Khalaf}(2018)}]{Khalaf2018}%
  \BibitemOpen
  \bibfield  {author} {\bibinfo {author} {\bibfnamefont {E.}~\bibnamefont
  {Khalaf}},\ }\bibfield  {title} {\enquote {\bibinfo {title} {Higher-order
  topological insulators and superconductors protected by inversion
  symmetry},}\ }\href {\doibase 10.1103/PhysRevB.97.205136} {\bibfield
  {journal} {\bibinfo  {journal} {Phys. Rev. B}\ }\textbf {\bibinfo {volume}
  {97}},\ \bibinfo {pages} {205136} (\bibinfo {year} {2018})}\BibitemShut
  {NoStop}%
\bibitem [{\citenamefont {Zhang}\ \emph {et~al.}(2019)\citenamefont {Zhang},
  \citenamefont {Cole}, \citenamefont {Wu},\ and\ \citenamefont
  {Das~Sarma}}]{ZhangPRL2019}%
  \BibitemOpen
  \bibfield  {author} {\bibinfo {author} {\bibfnamefont {R.-X.}\ \bibnamefont
  {Zhang}}, \bibinfo {author} {\bibfnamefont {W.~S.}\ \bibnamefont {Cole}},
  \bibinfo {author} {\bibfnamefont {X.}~\bibnamefont {Wu}}, \ and\ \bibinfo
  {author} {\bibfnamefont {S.}~\bibnamefont {Das~Sarma}},\ }\bibfield  {title}
  {\enquote {\bibinfo {title} {Higher-order topology and nodal topological
  superconductivity in fe(se,te) heterostructures},}\ }\href {\doibase
  10.1103/PhysRevLett.123.167001} {\bibfield  {journal} {\bibinfo  {journal}
  {Phys. Rev. Lett.}\ }\textbf {\bibinfo {volume} {123}},\ \bibinfo {pages}
  {167001} (\bibinfo {year} {2019})}\BibitemShut {NoStop}%
\bibitem [{\citenamefont {Hsu}\ \emph {et~al.}(2020)\citenamefont {Hsu},
  \citenamefont {Cole}, \citenamefont {Zhang},\ and\ \citenamefont
  {Sau}}]{HsuPRL2020}%
  \BibitemOpen
  \bibfield  {author} {\bibinfo {author} {\bibfnamefont {Y.-T.}\ \bibnamefont
  {Hsu}}, \bibinfo {author} {\bibfnamefont {W.~S.}\ \bibnamefont {Cole}},
  \bibinfo {author} {\bibfnamefont {R.-X.}\ \bibnamefont {Zhang}}, \ and\
  \bibinfo {author} {\bibfnamefont {J.~D.}\ \bibnamefont {Sau}},\ }\bibfield
  {title} {\enquote {\bibinfo {title} {Inversion-protected higher-order
  topological superconductivity in monolayer ${\mathrm{wte}}_{2}$},}\ }\href
  {\doibase 10.1103/PhysRevLett.125.097001} {\bibfield  {journal} {\bibinfo
  {journal} {Phys. Rev. Lett.}\ }\textbf {\bibinfo {volume} {125}},\ \bibinfo
  {pages} {097001} (\bibinfo {year} {2020})}\BibitemShut {NoStop}%
\bibitem [{\citenamefont {Huang}\ \emph {et~al.}(2024)\citenamefont {Huang},
  \citenamefont {Park},\ and\ \citenamefont {Hsu}}]{MoTe2_Hsu}%
  \BibitemOpen
  \bibfield  {author} {\bibinfo {author} {\bibfnamefont {S.-J.}\ \bibnamefont
  {Huang}}, \bibinfo {author} {\bibfnamefont {K.}~\bibnamefont {Park}}, \ and\
  \bibinfo {author} {\bibfnamefont {Y.-T.}\ \bibnamefont {Hsu}},\ }\bibfield
  {title} {\enquote {\bibinfo {title} {Hybrid-order topological
  superconductivity in a topological metal 1t'-mote2},}\ }\href {\doibase
  10.1038/s41535-024-00633-7} {\bibfield  {journal} {\bibinfo  {journal} {npj
  Quantum Materials}\ }\textbf {\bibinfo {volume} {9}},\ \bibinfo {pages} {21}
  (\bibinfo {year} {2024})}\BibitemShut {NoStop}%
\bibitem [{\citenamefont {Feldman}\ \emph {et~al.}(2017)\citenamefont
  {Feldman}, \citenamefont {Randeria}, \citenamefont {Li}, \citenamefont
  {Jeon}, \citenamefont {Xie}, \citenamefont {Wang}, \citenamefont {Drozdov},
  \citenamefont {Andrei~Bernevig},\ and\ \citenamefont
  {Yazdani}}]{Feldman:2017aa}%
  \BibitemOpen
  \bibfield  {author} {\bibinfo {author} {\bibfnamefont {B.~E.}\ \bibnamefont
  {Feldman}}, \bibinfo {author} {\bibfnamefont {M.~T.}\ \bibnamefont
  {Randeria}}, \bibinfo {author} {\bibfnamefont {J.}~\bibnamefont {Li}},
  \bibinfo {author} {\bibfnamefont {S.}~\bibnamefont {Jeon}}, \bibinfo {author}
  {\bibfnamefont {Y.}~\bibnamefont {Xie}}, \bibinfo {author} {\bibfnamefont
  {Z.}~\bibnamefont {Wang}}, \bibinfo {author} {\bibfnamefont {I.~K.}\
  \bibnamefont {Drozdov}}, \bibinfo {author} {\bibfnamefont {B.}~\bibnamefont
  {Andrei~Bernevig}}, \ and\ \bibinfo {author} {\bibfnamefont {A.}~\bibnamefont
  {Yazdani}},\ }\bibfield  {title} {\enquote {\bibinfo {title} {High-resolution
  studies of the majorana atomic chain platform},}\ }\href {\doibase
  10.1038/nphys3947} {\bibfield  {journal} {\bibinfo  {journal} {Nat. Phys.}\
  }\textbf {\bibinfo {volume} {13}},\ \bibinfo {pages} {286--291} (\bibinfo
  {year} {2017})}\BibitemShut {NoStop}%
\bibitem [{\citenamefont {M{\'e}nard}\ \emph {et~al.}(2017)\citenamefont
  {M{\'e}nard}, \citenamefont {Guissart}, \citenamefont {Brun}, \citenamefont
  {Leriche}, \citenamefont {Trif}, \citenamefont {Debontridder}, \citenamefont
  {Demaille}, \citenamefont {Roditchev}, \citenamefont {Simon},\ and\
  \citenamefont {Cren}}]{Exp_2DTscProx_Ncomm2017}%
  \BibitemOpen
  \bibfield  {author} {\bibinfo {author} {\bibfnamefont {G.~C.}\ \bibnamefont
  {M{\'e}nard}}, \bibinfo {author} {\bibfnamefont {S.}~\bibnamefont
  {Guissart}}, \bibinfo {author} {\bibfnamefont {C.}~\bibnamefont {Brun}},
  \bibinfo {author} {\bibfnamefont {R.~T.}\ \bibnamefont {Leriche}}, \bibinfo
  {author} {\bibfnamefont {M.}~\bibnamefont {Trif}}, \bibinfo {author}
  {\bibfnamefont {F.}~\bibnamefont {Debontridder}}, \bibinfo {author}
  {\bibfnamefont {D.}~\bibnamefont {Demaille}}, \bibinfo {author}
  {\bibfnamefont {D.}~\bibnamefont {Roditchev}}, \bibinfo {author}
  {\bibfnamefont {P.}~\bibnamefont {Simon}}, \ and\ \bibinfo {author}
  {\bibfnamefont {T.}~\bibnamefont {Cren}},\ }\bibfield  {title} {\enquote
  {\bibinfo {title} {Two-dimensional topological superconductivity in
  pb/co/si(111)},}\ }\href {\doibase 10.1038/s41467-017-02192-x} {\bibfield
  {journal} {\bibinfo  {journal} {Nat. Commun.}\ }\textbf {\bibinfo {volume}
  {8}},\ \bibinfo {pages} {2040} (\bibinfo {year} {2017})}\BibitemShut
  {NoStop}%
\bibitem [{\citenamefont {Zhang}\ \emph
  {et~al.}(2018{\natexlab{a}})\citenamefont {Zhang}, \citenamefont {Yaji},
  \citenamefont {Hashimoto}, \citenamefont {Ota}, \citenamefont {Kondo},
  \citenamefont {Okazaki}, \citenamefont {Wang}, \citenamefont {Wen},
  \citenamefont {Gu}, \citenamefont {Ding},\ and\ \citenamefont
  {Shin}}]{Exp_MajoFeSc_Peng2018}%
  \BibitemOpen
  \bibfield  {author} {\bibinfo {author} {\bibfnamefont {P.}~\bibnamefont
  {Zhang}}, \bibinfo {author} {\bibfnamefont {K.}~\bibnamefont {Yaji}},
  \bibinfo {author} {\bibfnamefont {T.}~\bibnamefont {Hashimoto}}, \bibinfo
  {author} {\bibfnamefont {Y.}~\bibnamefont {Ota}}, \bibinfo {author}
  {\bibfnamefont {T.}~\bibnamefont {Kondo}}, \bibinfo {author} {\bibfnamefont
  {K.}~\bibnamefont {Okazaki}}, \bibinfo {author} {\bibfnamefont
  {Z.}~\bibnamefont {Wang}}, \bibinfo {author} {\bibfnamefont {J.}~\bibnamefont
  {Wen}}, \bibinfo {author} {\bibfnamefont {G.~D.}\ \bibnamefont {Gu}},
  \bibinfo {author} {\bibfnamefont {H.}~\bibnamefont {Ding}}, \ and\ \bibinfo
  {author} {\bibfnamefont {S.}~\bibnamefont {Shin}},\ }\bibfield  {title}
  {\enquote {\bibinfo {title} {Observation of topological superconductivity on
  the surface of an iron-based superconductor},}\ }\href {\doibase
  10.1126/science.aan4596} {\bibfield  {journal} {\bibinfo  {journal}
  {Science}\ }\textbf {\bibinfo {volume} {360}},\ \bibinfo {pages} {182--186}
  (\bibinfo {year} {2018}{\natexlab{a}})}\BibitemShut {NoStop}%
\bibitem [{\citenamefont {Wang}\ \emph {et~al.}(2018)\citenamefont {Wang},
  \citenamefont {Kong}, \citenamefont {Fan}, \citenamefont {Chen},
  \citenamefont {Zhu}, \citenamefont {Liu}, \citenamefont {Cao}, \citenamefont
  {Sun}, \citenamefont {Du}, \citenamefont {Schneeloch}, \citenamefont {Zhong},
  \citenamefont {Gu}, \citenamefont {Fu}, \citenamefont {Ding},\ and\
  \citenamefont {Gao}}]{Exp_MajoFeSc_Wang2018}%
  \BibitemOpen
  \bibfield  {author} {\bibinfo {author} {\bibfnamefont {D.}~\bibnamefont
  {Wang}}, \bibinfo {author} {\bibfnamefont {L.}~\bibnamefont {Kong}}, \bibinfo
  {author} {\bibfnamefont {P.}~\bibnamefont {Fan}}, \bibinfo {author}
  {\bibfnamefont {H.}~\bibnamefont {Chen}}, \bibinfo {author} {\bibfnamefont
  {S.}~\bibnamefont {Zhu}}, \bibinfo {author} {\bibfnamefont {W.}~\bibnamefont
  {Liu}}, \bibinfo {author} {\bibfnamefont {L.}~\bibnamefont {Cao}}, \bibinfo
  {author} {\bibfnamefont {Y.}~\bibnamefont {Sun}}, \bibinfo {author}
  {\bibfnamefont {S.}~\bibnamefont {Du}}, \bibinfo {author} {\bibfnamefont
  {J.}~\bibnamefont {Schneeloch}}, \bibinfo {author} {\bibfnamefont
  {R.}~\bibnamefont {Zhong}}, \bibinfo {author} {\bibfnamefont
  {G.}~\bibnamefont {Gu}}, \bibinfo {author} {\bibfnamefont {L.}~\bibnamefont
  {Fu}}, \bibinfo {author} {\bibfnamefont {H.}~\bibnamefont {Ding}}, \ and\
  \bibinfo {author} {\bibfnamefont {H.-J.}\ \bibnamefont {Gao}},\ }\bibfield
  {title} {\enquote {\bibinfo {title} {Evidence for majorana bound states in an
  iron-based superconductor},}\ }\href {\doibase 10.1126/science.aao1797}
  {\bibfield  {journal} {\bibinfo  {journal} {Science}\ }\textbf {\bibinfo
  {volume} {362}},\ \bibinfo {pages} {333--335} (\bibinfo {year}
  {2018})}\BibitemShut {NoStop}%
\bibitem [{\citenamefont {J{\"a}ck}\ \emph {et~al.}(2019)\citenamefont
  {J{\"a}ck}, \citenamefont {Xie}, \citenamefont {Li}, \citenamefont {Jeon},
  \citenamefont {Bernevig},\ and\ \citenamefont
  {Yazdani}}]{Exp_MajoSTM2DTsc_Yazdani}%
  \BibitemOpen
  \bibfield  {author} {\bibinfo {author} {\bibfnamefont {B.}~\bibnamefont
  {J{\"a}ck}}, \bibinfo {author} {\bibfnamefont {Y.}~\bibnamefont {Xie}},
  \bibinfo {author} {\bibfnamefont {J.}~\bibnamefont {Li}}, \bibinfo {author}
  {\bibfnamefont {S.}~\bibnamefont {Jeon}}, \bibinfo {author} {\bibfnamefont
  {B.~A.}\ \bibnamefont {Bernevig}}, \ and\ \bibinfo {author} {\bibfnamefont
  {A.}~\bibnamefont {Yazdani}},\ }\bibfield  {title} {\enquote {\bibinfo
  {title} {Observation of a majorana zero mode in a topologically protected
  edge channel},}\ }\href {\doibase 10.1126/science.aax1444} {\bibfield
  {journal} {\bibinfo  {journal} {Science}\ }\textbf {\bibinfo {volume}
  {364}},\ \bibinfo {pages} {1255--1259} (\bibinfo {year} {2019})}\BibitemShut
  {NoStop}%
\bibitem [{\citenamefont {Palacio-Morales}\ \emph {et~al.}(2019)\citenamefont
  {Palacio-Morales}, \citenamefont {Mascot}, \citenamefont {Cocklin},
  \citenamefont {Kim}, \citenamefont {Rachel}, \citenamefont {Morr},\ and\
  \citenamefont {Wiesendanger}}]{Exp_STMMajoedge_2019}%
  \BibitemOpen
  \bibfield  {author} {\bibinfo {author} {\bibfnamefont {A.}~\bibnamefont
  {Palacio-Morales}}, \bibinfo {author} {\bibfnamefont {E.}~\bibnamefont
  {Mascot}}, \bibinfo {author} {\bibfnamefont {S.}~\bibnamefont {Cocklin}},
  \bibinfo {author} {\bibfnamefont {H.}~\bibnamefont {Kim}}, \bibinfo {author}
  {\bibfnamefont {S.}~\bibnamefont {Rachel}}, \bibinfo {author} {\bibfnamefont
  {D.~K.}\ \bibnamefont {Morr}}, \ and\ \bibinfo {author} {\bibfnamefont
  {R.}~\bibnamefont {Wiesendanger}},\ }\bibfield  {title} {\enquote {\bibinfo
  {title} {Atomic-scale interface engineering of majorana edge modes in a 2d
  magnet-superconductor hybrid system},}\ }\href {\doibase
  10.1126/sciadv.aav6600} {\bibfield  {journal} {\bibinfo  {journal} {Science
  Advances}\ }\textbf {\bibinfo {volume} {5}},\ \bibinfo {pages} {eaav6600}
  (\bibinfo {year} {2019})}\BibitemShut {NoStop}%
\bibitem [{\citenamefont {Manna}\ \emph {et~al.}(2020)\citenamefont {Manna},
  \citenamefont {Wei}, \citenamefont {Xie}, \citenamefont {Law}, \citenamefont
  {Lee},\ and\ \citenamefont {Moodera}}]{Exp_gold_Manna2020}%
  \BibitemOpen
  \bibfield  {author} {\bibinfo {author} {\bibfnamefont {S.}~\bibnamefont
  {Manna}}, \bibinfo {author} {\bibfnamefont {P.}~\bibnamefont {Wei}}, \bibinfo
  {author} {\bibfnamefont {Y.}~\bibnamefont {Xie}}, \bibinfo {author}
  {\bibfnamefont {K.~T.}\ \bibnamefont {Law}}, \bibinfo {author} {\bibfnamefont
  {P.~A.}\ \bibnamefont {Lee}}, \ and\ \bibinfo {author} {\bibfnamefont
  {J.~S.}\ \bibnamefont {Moodera}},\ }\bibfield  {title} {\enquote {\bibinfo
  {title} {Signature of a pair of majorana zero modes in superconducting gold
  surface states},}\ }\href {\doibase 10.1073/pnas.1919753117} {\bibfield
  {journal} {\bibinfo  {journal} {Proc. Natl. Acad. Sci. U.S.A.}\ }\textbf
  {\bibinfo {volume} {117}},\ \bibinfo {pages} {8775--8782} (\bibinfo {year}
  {2020})}\BibitemShut {NoStop}%
\bibitem [{\citenamefont {Kezilebieke}\ \emph {et~al.}(2020)\citenamefont
  {Kezilebieke}, \citenamefont {Huda}, \citenamefont {Va{\v n}o}, \citenamefont
  {Aapro}, \citenamefont {Ganguli}, \citenamefont {Silveira}, \citenamefont
  {G{\l}odzik}, \citenamefont {Foster}, \citenamefont {Ojanen},\ and\
  \citenamefont {Liljeroth}}]{Exp_VdW_Kezilebieke2020}%
  \BibitemOpen
  \bibfield  {author} {\bibinfo {author} {\bibfnamefont {S.}~\bibnamefont
  {Kezilebieke}}, \bibinfo {author} {\bibfnamefont {M.~N.}\ \bibnamefont
  {Huda}}, \bibinfo {author} {\bibfnamefont {V.}~\bibnamefont {Va{\v n}o}},
  \bibinfo {author} {\bibfnamefont {M.}~\bibnamefont {Aapro}}, \bibinfo
  {author} {\bibfnamefont {S.~C.}\ \bibnamefont {Ganguli}}, \bibinfo {author}
  {\bibfnamefont {O.~J.}\ \bibnamefont {Silveira}}, \bibinfo {author}
  {\bibfnamefont {S.}~\bibnamefont {G{\l}odzik}}, \bibinfo {author}
  {\bibfnamefont {A.~S.}\ \bibnamefont {Foster}}, \bibinfo {author}
  {\bibfnamefont {T.}~\bibnamefont {Ojanen}}, \ and\ \bibinfo {author}
  {\bibfnamefont {P.}~\bibnamefont {Liljeroth}},\ }\bibfield  {title} {\enquote
  {\bibinfo {title} {Topological superconductivity in a van der waals
  heterostructure},}\ }\href {\doibase 10.1038/s41586-020-2989-y} {\bibfield
  {journal} {\bibinfo  {journal} {Nature}\ }\textbf {\bibinfo {volume} {588}},\
  \bibinfo {pages} {424--428} (\bibinfo {year} {2020})}\BibitemShut {NoStop}%
\bibitem [{\citenamefont {Wang}\ \emph {et~al.}(2020)\citenamefont {Wang},
  \citenamefont {Rodriguez}, \citenamefont {Jiao}, \citenamefont {Howard},
  \citenamefont {Graham}, \citenamefont {Gu}, \citenamefont {Hughes},
  \citenamefont {Morr},\ and\ \citenamefont
  {Madhavan}}]{Exp_FeSc_MajoedgeSTM_2020}%
  \BibitemOpen
  \bibfield  {author} {\bibinfo {author} {\bibfnamefont {Z.}~\bibnamefont
  {Wang}}, \bibinfo {author} {\bibfnamefont {J.~O.}\ \bibnamefont {Rodriguez}},
  \bibinfo {author} {\bibfnamefont {L.}~\bibnamefont {Jiao}}, \bibinfo {author}
  {\bibfnamefont {S.}~\bibnamefont {Howard}}, \bibinfo {author} {\bibfnamefont
  {M.}~\bibnamefont {Graham}}, \bibinfo {author} {\bibfnamefont {G.~D.}\
  \bibnamefont {Gu}}, \bibinfo {author} {\bibfnamefont {T.~L.}\ \bibnamefont
  {Hughes}}, \bibinfo {author} {\bibfnamefont {D.~K.}\ \bibnamefont {Morr}}, \
  and\ \bibinfo {author} {\bibfnamefont {V.}~\bibnamefont {Madhavan}},\
  }\bibfield  {title} {\enquote {\bibinfo {title} {Evidence for dispersing 1d
  majorana channels in an iron-based superconductor},}\ }\href {\doibase
  10.1126/science.aaw8419} {\bibfield  {journal} {\bibinfo  {journal}
  {Science}\ }\textbf {\bibinfo {volume} {367}},\ \bibinfo {pages} {104--108}
  (\bibinfo {year} {2020})}\BibitemShut {NoStop}%
\bibitem [{\citenamefont {J{\"a}ck}\ \emph {et~al.}(2021)\citenamefont
  {J{\"a}ck}, \citenamefont {Xie},\ and\ \citenamefont
  {Yazdani}}]{STMreview_Majo}%
  \BibitemOpen
  \bibfield  {author} {\bibinfo {author} {\bibfnamefont {B.}~\bibnamefont
  {J{\"a}ck}}, \bibinfo {author} {\bibfnamefont {Y.}~\bibnamefont {Xie}}, \
  and\ \bibinfo {author} {\bibfnamefont {A.}~\bibnamefont {Yazdani}},\
  }\bibfield  {title} {\enquote {\bibinfo {title} {Detecting and distinguishing
  majorana zero modes with the scanning tunnelling microscope},}\ }\href
  {\doibase 10.1038/s42254-021-00328-z} {\bibfield  {journal} {\bibinfo
  {journal} {Nat. Rev. Phys.}\ }\textbf {\bibinfo {volume} {3}},\ \bibinfo
  {pages} {541--554} (\bibinfo {year} {2021})}\BibitemShut {NoStop}%
\bibitem [{\citenamefont {Yu}\ \emph {et~al.}(2021)\citenamefont {Yu},
  \citenamefont {Chen}, \citenamefont {Gomanko}, \citenamefont {Badawy},
  \citenamefont {Bakkers}, \citenamefont {Zuo}, \citenamefont {Mourik},\ and\
  \citenamefont {Frolov}}]{Exp_Nonmajo_Frolov}%
  \BibitemOpen
  \bibfield  {author} {\bibinfo {author} {\bibfnamefont {P.}~\bibnamefont
  {Yu}}, \bibinfo {author} {\bibfnamefont {J.}~\bibnamefont {Chen}}, \bibinfo
  {author} {\bibfnamefont {M.}~\bibnamefont {Gomanko}}, \bibinfo {author}
  {\bibfnamefont {G.}~\bibnamefont {Badawy}}, \bibinfo {author} {\bibfnamefont
  {E.~P. A.~M.}\ \bibnamefont {Bakkers}}, \bibinfo {author} {\bibfnamefont
  {K.}~\bibnamefont {Zuo}}, \bibinfo {author} {\bibfnamefont {V.}~\bibnamefont
  {Mourik}}, \ and\ \bibinfo {author} {\bibfnamefont {S.~M.}\ \bibnamefont
  {Frolov}},\ }\bibfield  {title} {\enquote {\bibinfo {title} {Non-majorana
  states yield nearly quantized conductance in proximatized nanowires},}\
  }\href {\doibase 10.1038/s41567-020-01107-w} {\bibfield  {journal} {\bibinfo
  {journal} {Nat. Phys.}\ }\textbf {\bibinfo {volume} {17}},\ \bibinfo {pages}
  {482--488} (\bibinfo {year} {2021})}\BibitemShut {NoStop}%
\bibitem [{\citenamefont {Peng}\ \emph {et~al.}(2015)\citenamefont {Peng},
  \citenamefont {Pientka}, \citenamefont {Vinkler-Aviv}, \citenamefont
  {Glazman},\ and\ \citenamefont {von Oppen}}]{Nonmajo_VonOppen_2015}%
  \BibitemOpen
  \bibfield  {author} {\bibinfo {author} {\bibfnamefont {Y.}~\bibnamefont
  {Peng}}, \bibinfo {author} {\bibfnamefont {F.}~\bibnamefont {Pientka}},
  \bibinfo {author} {\bibfnamefont {Y.}~\bibnamefont {Vinkler-Aviv}}, \bibinfo
  {author} {\bibfnamefont {L.~I.}\ \bibnamefont {Glazman}}, \ and\ \bibinfo
  {author} {\bibfnamefont {F.}~\bibnamefont {von Oppen}},\ }\bibfield  {title}
  {\enquote {\bibinfo {title} {Robust majorana conductance peaks for a
  superconducting lead},}\ }\href {\doibase 10.1103/PhysRevLett.115.266804}
  {\bibfield  {journal} {\bibinfo  {journal} {Phys. Rev. Lett.}\ }\textbf
  {\bibinfo {volume} {115}},\ \bibinfo {pages} {266804} (\bibinfo {year}
  {2015})}\BibitemShut {NoStop}%
\bibitem [{\citenamefont {Li}\ \emph {et~al.}(2018)\citenamefont {Li},
  \citenamefont {Jeon}, \citenamefont {Xie}, \citenamefont {Yazdani},\ and\
  \citenamefont {Bernevig}}]{MajoSpin_Bernevig2018}%
  \BibitemOpen
  \bibfield  {author} {\bibinfo {author} {\bibfnamefont {J.}~\bibnamefont
  {Li}}, \bibinfo {author} {\bibfnamefont {S.}~\bibnamefont {Jeon}}, \bibinfo
  {author} {\bibfnamefont {Y.}~\bibnamefont {Xie}}, \bibinfo {author}
  {\bibfnamefont {A.}~\bibnamefont {Yazdani}}, \ and\ \bibinfo {author}
  {\bibfnamefont {B.~A.}\ \bibnamefont {Bernevig}},\ }\bibfield  {title}
  {\enquote {\bibinfo {title} {Majorana spin in magnetic atomic chain
  systems},}\ }\href {\doibase 10.1103/PhysRevB.97.125119} {\bibfield
  {journal} {\bibinfo  {journal} {Phys. Rev. B}\ }\textbf {\bibinfo {volume}
  {97}},\ \bibinfo {pages} {125119} (\bibinfo {year} {2018})}\BibitemShut
  {NoStop}%
\bibitem [{\citenamefont {Zhang}\ \emph
  {et~al.}(2018{\natexlab{b}})\citenamefont {Zhang}, \citenamefont {Liu},
  \citenamefont {Gazibegovic}, \citenamefont {Xu}, \citenamefont {Logan},
  \citenamefont {Wang}, \citenamefont {van Loo}, \citenamefont {Bommer},
  \citenamefont {de~Moor}, \citenamefont {Car}, \citenamefont {Op~het Veld},
  \citenamefont {van Veldhoven}, \citenamefont {Koelling}, \citenamefont
  {Verheijen}, \citenamefont {Pendharkar}, \citenamefont {Pennachio},
  \citenamefont {Shojaei}, \citenamefont {Lee}, \citenamefont {Palmstr{\o}m},
  \citenamefont {Bakkers}, \citenamefont {Sarma},\ and\ \citenamefont
  {Kouwenhoven}}]{RetractMajo2018}%
  \BibitemOpen
  \bibfield  {author} {\bibinfo {author} {\bibfnamefont {H.}~\bibnamefont
  {Zhang}}, \bibinfo {author} {\bibfnamefont {C.-X.}\ \bibnamefont {Liu}},
  \bibinfo {author} {\bibfnamefont {S.}~\bibnamefont {Gazibegovic}}, \bibinfo
  {author} {\bibfnamefont {D.}~\bibnamefont {Xu}}, \bibinfo {author}
  {\bibfnamefont {J.~A.}\ \bibnamefont {Logan}}, \bibinfo {author}
  {\bibfnamefont {G.}~\bibnamefont {Wang}}, \bibinfo {author} {\bibfnamefont
  {N.}~\bibnamefont {van Loo}}, \bibinfo {author} {\bibfnamefont {J.~D.~S.}\
  \bibnamefont {Bommer}}, \bibinfo {author} {\bibfnamefont {M.~W.~A.}\
  \bibnamefont {de~Moor}}, \bibinfo {author} {\bibfnamefont {D.}~\bibnamefont
  {Car}}, \bibinfo {author} {\bibfnamefont {R.~L.~M.}\ \bibnamefont {Op~het
  Veld}}, \bibinfo {author} {\bibfnamefont {P.~J.}\ \bibnamefont {van
  Veldhoven}}, \bibinfo {author} {\bibfnamefont {S.}~\bibnamefont {Koelling}},
  \bibinfo {author} {\bibfnamefont {M.~A.}\ \bibnamefont {Verheijen}}, \bibinfo
  {author} {\bibfnamefont {M.}~\bibnamefont {Pendharkar}}, \bibinfo {author}
  {\bibfnamefont {D.~J.}\ \bibnamefont {Pennachio}}, \bibinfo {author}
  {\bibfnamefont {B.}~\bibnamefont {Shojaei}}, \bibinfo {author} {\bibfnamefont
  {J.~S.}\ \bibnamefont {Lee}}, \bibinfo {author} {\bibfnamefont {C.~J.}\
  \bibnamefont {Palmstr{\o}m}}, \bibinfo {author} {\bibfnamefont {E.~P. A.~M.}\
  \bibnamefont {Bakkers}}, \bibinfo {author} {\bibfnamefont {S.~D.}\
  \bibnamefont {Sarma}}, \ and\ \bibinfo {author} {\bibfnamefont {L.~P.}\
  \bibnamefont {Kouwenhoven}},\ }\bibfield  {title} {\enquote {\bibinfo {title}
  {Retracted article: Quantized majorana conductance},}\ }\href {\doibase
  10.1038/nature26142} {\bibfield  {journal} {\bibinfo  {journal} {Nature}\
  }\textbf {\bibinfo {volume} {556}},\ \bibinfo {pages} {74--79} (\bibinfo
  {year} {2018}{\natexlab{b}})}\BibitemShut {NoStop}%
\bibitem [{\citenamefont {Pan}\ and\ \citenamefont
  {Das~Sarma}(2020)}]{Thy_ZBP_Sankar}%
  \BibitemOpen
  \bibfield  {author} {\bibinfo {author} {\bibfnamefont {H.}~\bibnamefont
  {Pan}}\ and\ \bibinfo {author} {\bibfnamefont {S.}~\bibnamefont
  {Das~Sarma}},\ }\bibfield  {title} {\enquote {\bibinfo {title} {Physical
  mechanisms for zero-bias conductance peaks in majorana nanowires},}\ }\href
  {\doibase 10.1103/PhysRevResearch.2.013377} {\bibfield  {journal} {\bibinfo
  {journal} {Phys. Rev. Res.}\ }\textbf {\bibinfo {volume} {2}},\ \bibinfo
  {pages} {013377} (\bibinfo {year} {2020})}\BibitemShut {NoStop}%
\bibitem [{\citenamefont {Frolov}\ \emph {et~al.}(2020)\citenamefont {Frolov},
  \citenamefont {Manfra},\ and\ \citenamefont {Sau}}]{Review_nonmajo_Sau}%
  \BibitemOpen
  \bibfield  {author} {\bibinfo {author} {\bibfnamefont {S.~M.}\ \bibnamefont
  {Frolov}}, \bibinfo {author} {\bibfnamefont {M.~J.}\ \bibnamefont {Manfra}},
  \ and\ \bibinfo {author} {\bibfnamefont {J.~D.}\ \bibnamefont {Sau}},\
  }\bibfield  {title} {\enquote {\bibinfo {title} {Topological
  superconductivity in hybrid devices},}\ }\href {\doibase
  10.1038/s41567-020-0925-6} {\bibfield  {journal} {\bibinfo  {journal} {Nat.
  Phys.}\ }\textbf {\bibinfo {volume} {16}},\ \bibinfo {pages} {718--724}
  (\bibinfo {year} {2020})}\BibitemShut {NoStop}%
\bibitem [{\citenamefont {Pan}\ \emph {et~al.}(2020)\citenamefont {Pan},
  \citenamefont {Cole}, \citenamefont {Sau},\ and\ \citenamefont
  {Das~Sarma}}]{Thy_nonMajo_Sankar_PRB2020}%
  \BibitemOpen
  \bibfield  {author} {\bibinfo {author} {\bibfnamefont {H.}~\bibnamefont
  {Pan}}, \bibinfo {author} {\bibfnamefont {W.~S.}\ \bibnamefont {Cole}},
  \bibinfo {author} {\bibfnamefont {J.~D.}\ \bibnamefont {Sau}}, \ and\
  \bibinfo {author} {\bibfnamefont {S.}~\bibnamefont {Das~Sarma}},\ }\bibfield
  {title} {\enquote {\bibinfo {title} {Generic quantized zero-bias conductance
  peaks in superconductor-semiconductor hybrid structures},}\ }\href {\doibase
  10.1103/PhysRevB.101.024506} {\bibfield  {journal} {\bibinfo  {journal}
  {Phys. Rev. B}\ }\textbf {\bibinfo {volume} {101}},\ \bibinfo {pages}
  {024506} (\bibinfo {year} {2020})}\BibitemShut {NoStop}%
\bibitem [{\citenamefont {Valentini}\ \emph {et~al.}(2021)\citenamefont
  {Valentini}, \citenamefont {Pe{\~n}aranda}, \citenamefont {Hofmann},
  \citenamefont {Brauns}, \citenamefont {Hauschild}, \citenamefont {Krogstrup},
  \citenamefont {San-Jose}, \citenamefont {Prada}, \citenamefont {Aguado},\
  and\ \citenamefont {Katsaros}}]{Exp_nonMajo_Katsaros2021}%
  \BibitemOpen
  \bibfield  {author} {\bibinfo {author} {\bibfnamefont {M.}~\bibnamefont
  {Valentini}}, \bibinfo {author} {\bibfnamefont {F.}~\bibnamefont
  {Pe{\~n}aranda}}, \bibinfo {author} {\bibfnamefont {A.}~\bibnamefont
  {Hofmann}}, \bibinfo {author} {\bibfnamefont {M.}~\bibnamefont {Brauns}},
  \bibinfo {author} {\bibfnamefont {R.}~\bibnamefont {Hauschild}}, \bibinfo
  {author} {\bibfnamefont {P.}~\bibnamefont {Krogstrup}}, \bibinfo {author}
  {\bibfnamefont {P.}~\bibnamefont {San-Jose}}, \bibinfo {author}
  {\bibfnamefont {E.}~\bibnamefont {Prada}}, \bibinfo {author} {\bibfnamefont
  {R.}~\bibnamefont {Aguado}}, \ and\ \bibinfo {author} {\bibfnamefont
  {G.}~\bibnamefont {Katsaros}},\ }\bibfield  {title} {\enquote {\bibinfo
  {title} {Nontopological zero-bias peaks in full-shell nanowires induced by
  flux-tunable andreev states},}\ }\href {\doibase 10.1126/science.abf1513}
  {\bibfield  {journal} {\bibinfo  {journal} {Science}\ }\textbf {\bibinfo
  {volume} {373}},\ \bibinfo {pages} {82--88} (\bibinfo {year}
  {2021})}\BibitemShut {NoStop}%
\bibitem [{\citenamefont {Sundaram}\ and\ \citenamefont {Niu}(1999)}]{Niu1999}%
  \BibitemOpen
  \bibfield  {author} {\bibinfo {author} {\bibfnamefont {G.}~\bibnamefont
  {Sundaram}}\ and\ \bibinfo {author} {\bibfnamefont {Q.}~\bibnamefont {Niu}},\
  }\bibfield  {title} {\enquote {\bibinfo {title} {Wave-packet dynamics in
  slowly perturbed crystals: Gradient corrections and berry-phase effects},}\
  }\href@noop {} {\bibfield  {journal} {\bibinfo  {journal} {Phys. Rev. B}\
  }\textbf {\bibinfo {volume} {59}},\ \bibinfo {pages} {14915--14925} (\bibinfo
  {year} {1999})}\BibitemShut {NoStop}%
\bibitem [{\citenamefont {Xiao}\ \emph {et~al.}(2005)\citenamefont {Xiao},
  \citenamefont {Shi},\ and\ \citenamefont {Niu}}]{NiuDOS}%
  \BibitemOpen
  \bibfield  {author} {\bibinfo {author} {\bibfnamefont {D.}~\bibnamefont
  {Xiao}}, \bibinfo {author} {\bibfnamefont {J.}~\bibnamefont {Shi}}, \ and\
  \bibinfo {author} {\bibfnamefont {Q.}~\bibnamefont {Niu}},\ }\bibfield
  {title} {\enquote {\bibinfo {title} {Berry phase correction to electron
  density of states in solids},}\ }\href@noop {} {\bibfield  {journal}
  {\bibinfo  {journal} {Phys. Rev. Lett.}\ }\textbf {\bibinfo {volume} {95}},\
  \bibinfo {pages} {137204} (\bibinfo {year} {2005})}\BibitemShut {NoStop}%
\bibitem [{\citenamefont {Shindou}\ and\ \citenamefont
  {Imura}(2005)}]{Shindou}%
  \BibitemOpen
  \bibfield  {author} {\bibinfo {author} {\bibfnamefont {R.}~\bibnamefont
  {Shindou}}\ and\ \bibinfo {author} {\bibfnamefont {K.-I.}\ \bibnamefont
  {Imura}},\ }\bibfield  {title} {\enquote {\bibinfo {title} {Noncommutative
  geometry and non-abelian berry phase in the wave-packet dynamics of bloch
  electrons},}\ }\href@noop {} {\bibfield  {journal} {\bibinfo  {journal}
  {Nuclear Physics B}\ }\textbf {\bibinfo {volume} {720}},\ \bibinfo {pages}
  {399--435} (\bibinfo {year} {2005})}\BibitemShut {NoStop}%
\bibitem [{\citenamefont {Xiao}\ \emph {et~al.}(2010)\citenamefont {Xiao},
  \citenamefont {Chang},\ and\ \citenamefont {Niu}}]{NiuRev}%
  \BibitemOpen
  \bibfield  {author} {\bibinfo {author} {\bibfnamefont {D.}~\bibnamefont
  {Xiao}}, \bibinfo {author} {\bibfnamefont {M.-C.}\ \bibnamefont {Chang}}, \
  and\ \bibinfo {author} {\bibfnamefont {Q.}~\bibnamefont {Niu}},\ }\bibfield
  {title} {\enquote {\bibinfo {title} {Berry phase effects on electronic
  properties},}\ }\href@noop {} {\bibfield  {journal} {\bibinfo  {journal}
  {Rev. Mod. Phys.}\ }\textbf {\bibinfo {volume} {82}},\ \bibinfo {pages}
  {1959--2007} (\bibinfo {year} {2010})}\BibitemShut {NoStop}%
\bibitem [{\citenamefont {Gao}\ \emph {et~al.}(2015)\citenamefont {Gao},
  \citenamefont {Yang},\ and\ \citenamefont {Niu}}]{GaoPRB2015}%
  \BibitemOpen
  \bibfield  {author} {\bibinfo {author} {\bibfnamefont {Y.}~\bibnamefont
  {Gao}}, \bibinfo {author} {\bibfnamefont {S.~A.}\ \bibnamefont {Yang}}, \
  and\ \bibinfo {author} {\bibfnamefont {Q.}~\bibnamefont {Niu}},\ }\bibfield
  {title} {\enquote {\bibinfo {title} {Geometrical effects in orbital magnetic
  susceptibility},}\ }\href {\doibase 10.1103/PhysRevB.91.214405} {\bibfield
  {journal} {\bibinfo  {journal} {Phys. Rev. B}\ }\textbf {\bibinfo {volume}
  {91}},\ \bibinfo {pages} {214405} (\bibinfo {year} {2015})}\BibitemShut
  {NoStop}%
\bibitem [{\citenamefont {Julku}\ \emph {et~al.}(2016)\citenamefont {Julku},
  \citenamefont {Peotta}, \citenamefont {Vanhala}, \citenamefont {Kim},\ and\
  \citenamefont {T\"orm\"a}}]{JulkuPRL2016}%
  \BibitemOpen
  \bibfield  {author} {\bibinfo {author} {\bibfnamefont {A.}~\bibnamefont
  {Julku}}, \bibinfo {author} {\bibfnamefont {S.}~\bibnamefont {Peotta}},
  \bibinfo {author} {\bibfnamefont {T.~I.}\ \bibnamefont {Vanhala}}, \bibinfo
  {author} {\bibfnamefont {D.-H.}\ \bibnamefont {Kim}}, \ and\ \bibinfo
  {author} {\bibfnamefont {P.}~\bibnamefont {T\"orm\"a}},\ }\bibfield  {title}
  {\enquote {\bibinfo {title} {Geometric origin of superfluidity in the
  lieb-lattice flat band},}\ }\href {\doibase 10.1103/PhysRevLett.117.045303}
  {\bibfield  {journal} {\bibinfo  {journal} {Phys. Rev. Lett.}\ }\textbf
  {\bibinfo {volume} {117}},\ \bibinfo {pages} {045303} (\bibinfo {year}
  {2016})}\BibitemShut {NoStop}%
\bibitem [{\citenamefont {Liang}\ \emph
  {et~al.}(2017{\natexlab{a}})\citenamefont {Liang}, \citenamefont {Peotta},
  \citenamefont {Harju},\ and\ \citenamefont {T\"orm\"a}}]{LiangPRB2017}%
  \BibitemOpen
  \bibfield  {author} {\bibinfo {author} {\bibfnamefont {L.}~\bibnamefont
  {Liang}}, \bibinfo {author} {\bibfnamefont {S.}~\bibnamefont {Peotta}},
  \bibinfo {author} {\bibfnamefont {A.}~\bibnamefont {Harju}}, \ and\ \bibinfo
  {author} {\bibfnamefont {P.}~\bibnamefont {T\"orm\"a}},\ }\bibfield  {title}
  {\enquote {\bibinfo {title} {Wave-packet dynamics of bogoliubov
  quasiparticles: Quantum metric effects},}\ }\href {\doibase
  10.1103/PhysRevB.96.064511} {\bibfield  {journal} {\bibinfo  {journal} {Phys.
  Rev. B}\ }\textbf {\bibinfo {volume} {96}},\ \bibinfo {pages} {064511}
  (\bibinfo {year} {2017}{\natexlab{a}})}\BibitemShut {NoStop}%
\bibitem [{\citenamefont {Liang}\ \emph
  {et~al.}(2017{\natexlab{b}})\citenamefont {Liang}, \citenamefont {Vanhala},
  \citenamefont {Peotta}, \citenamefont {Siro}, \citenamefont {Harju},\ and\
  \citenamefont {T\"orm\"a}}]{LiangBCPRB2017}%
  \BibitemOpen
  \bibfield  {author} {\bibinfo {author} {\bibfnamefont {L.}~\bibnamefont
  {Liang}}, \bibinfo {author} {\bibfnamefont {T.~I.}\ \bibnamefont {Vanhala}},
  \bibinfo {author} {\bibfnamefont {S.}~\bibnamefont {Peotta}}, \bibinfo
  {author} {\bibfnamefont {T.}~\bibnamefont {Siro}}, \bibinfo {author}
  {\bibfnamefont {A.}~\bibnamefont {Harju}}, \ and\ \bibinfo {author}
  {\bibfnamefont {P.}~\bibnamefont {T\"orm\"a}},\ }\bibfield  {title} {\enquote
  {\bibinfo {title} {Band geometry, berry curvature, and superfluid weight},}\
  }\href {\doibase 10.1103/PhysRevB.95.024515} {\bibfield  {journal} {\bibinfo
  {journal} {Phys. Rev. B}\ }\textbf {\bibinfo {volume} {95}},\ \bibinfo
  {pages} {024515} (\bibinfo {year} {2017}{\natexlab{b}})}\BibitemShut
  {NoStop}%
\bibitem [{\citenamefont {Wang}\ \emph {et~al.}(2021)\citenamefont {Wang},
  \citenamefont {Dong}, \citenamefont {Xiao},\ and\ \citenamefont
  {Niu}}]{NiuSC}%
  \BibitemOpen
  \bibfield  {author} {\bibinfo {author} {\bibfnamefont {Z.}~\bibnamefont
  {Wang}}, \bibinfo {author} {\bibfnamefont {L.}~\bibnamefont {Dong}}, \bibinfo
  {author} {\bibfnamefont {C.}~\bibnamefont {Xiao}}, \ and\ \bibinfo {author}
  {\bibfnamefont {Q.}~\bibnamefont {Niu}},\ }\bibfield  {title} {\enquote
  {\bibinfo {title} {Berry curvature effects on quasiparticle dynamics in
  superconductors},}\ }\href@noop {} {\bibfield  {journal} {\bibinfo  {journal}
  {Phys. Rev. Lett.}\ }\textbf {\bibinfo {volume} {126}},\ \bibinfo {pages}
  {187001} (\bibinfo {year} {2021})}\BibitemShut {NoStop}%
\bibitem [{\citenamefont {Rossi}(2021)}]{RossiReivew2021}%
  \BibitemOpen
  \bibfield  {author} {\bibinfo {author} {\bibfnamefont {E.}~\bibnamefont
  {Rossi}},\ }\bibfield  {title} {\enquote {\bibinfo {title} {Quantum metric
  and correlated states in two-dimensional systems},}\ }\href {\doibase
  https://doi.org/10.1016/j.cossms.2021.100952} {\bibfield  {journal} {\bibinfo
   {journal} {Curr. Opin. Solid State Mater. Sci.}\ }\textbf {\bibinfo {volume}
  {25}},\ \bibinfo {pages} {100952} (\bibinfo {year} {2021})}\BibitemShut
  {NoStop}%
\bibitem [{\citenamefont {Ahn}\ and\ \citenamefont
  {Nagaosa}(2021)}]{AhnPRB2021}%
  \BibitemOpen
  \bibfield  {author} {\bibinfo {author} {\bibfnamefont {J.}~\bibnamefont
  {Ahn}}\ and\ \bibinfo {author} {\bibfnamefont {N.}~\bibnamefont {Nagaosa}},\
  }\bibfield  {title} {\enquote {\bibinfo {title} {Superconductivity-induced
  spectral weight transfer due to quantum geometry},}\ }\href {\doibase
  10.1103/PhysRevB.104.L100501} {\bibfield  {journal} {\bibinfo  {journal}
  {Phys. Rev. B}\ }\textbf {\bibinfo {volume} {104}},\ \bibinfo {pages}
  {L100501} (\bibinfo {year} {2021})}\BibitemShut {NoStop}%
\bibitem [{\citenamefont {Yu}\ \emph {et~al.}(2024)\citenamefont {Yu},
  \citenamefont {Ciccarino}, \citenamefont {Bianco}, \citenamefont {Errea},
  \citenamefont {Narang},\ and\ \citenamefont {Bernevig}}]{YuQG2023}%
  \BibitemOpen
  \bibfield  {author} {\bibinfo {author} {\bibfnamefont {J.}~\bibnamefont
  {Yu}}, \bibinfo {author} {\bibfnamefont {C.~J.}\ \bibnamefont {Ciccarino}},
  \bibinfo {author} {\bibfnamefont {R.}~\bibnamefont {Bianco}}, \bibinfo
  {author} {\bibfnamefont {I.}~\bibnamefont {Errea}}, \bibinfo {author}
  {\bibfnamefont {P.}~\bibnamefont {Narang}}, \ and\ \bibinfo {author}
  {\bibfnamefont {B.~A.}\ \bibnamefont {Bernevig}},\ }\bibfield  {title}
  {\enquote {\bibinfo {title} {Non-trivial quantum geometry and the strength of
  electron--phonon coupling},}\ }\href@noop {} {\bibfield  {journal} {\bibinfo
  {journal} {Nat. Phys.}\ ,\ \bibinfo {pages} {1--7}} (\bibinfo {year}
  {2024})}\BibitemShut {NoStop}%
\bibitem [{\citenamefont {Fu}\ and\ \citenamefont {Kane}(2008)}]{FuKane_TITsc}%
  \BibitemOpen
  \bibfield  {author} {\bibinfo {author} {\bibfnamefont {L.}~\bibnamefont
  {Fu}}\ and\ \bibinfo {author} {\bibfnamefont {C.~L.}\ \bibnamefont {Kane}},\
  }\bibfield  {title} {\enquote {\bibinfo {title} {Superconducting proximity
  effect and majorana fermions at the surface of a topological insulator},}\
  }\href {\doibase 10.1103/PhysRevLett.100.096407} {\bibfield  {journal}
  {\bibinfo  {journal} {Phys. Rev. Lett.}\ }\textbf {\bibinfo {volume} {100}},\
  \bibinfo {pages} {096407} (\bibinfo {year} {2008})}\BibitemShut {NoStop}%
\bibitem [{Ras()}]{RashbaTsc_Lee2009}%
  \BibitemOpen
  \href@noop {} {}\bibinfo {note} {P. A. Lee, arXiv e-prints, arXiv:0907.2681
  (2009).}\BibitemShut {Stop}%
\bibitem [{\citenamefont {Sau}\ \emph {et~al.}(2010{\natexlab{a}})\citenamefont
  {Sau}, \citenamefont {Lutchyn}, \citenamefont {Tewari},\ and\ \citenamefont
  {Das~Sarma}}]{SauPRL2010}%
  \BibitemOpen
  \bibfield  {author} {\bibinfo {author} {\bibfnamefont {J.~D.}\ \bibnamefont
  {Sau}}, \bibinfo {author} {\bibfnamefont {R.~M.}\ \bibnamefont {Lutchyn}},
  \bibinfo {author} {\bibfnamefont {S.}~\bibnamefont {Tewari}}, \ and\ \bibinfo
  {author} {\bibfnamefont {S.}~\bibnamefont {Das~Sarma}},\ }\bibfield  {title}
  {\enquote {\bibinfo {title} {Generic new platform for topological quantum
  computation using semiconductor heterostructures},}\ }\href {\doibase
  10.1103/PhysRevLett.104.040502} {\bibfield  {journal} {\bibinfo  {journal}
  {Phys. Rev. Lett.}\ }\textbf {\bibinfo {volume} {104}},\ \bibinfo {pages}
  {040502} (\bibinfo {year} {2010}{\natexlab{a}})}\BibitemShut {NoStop}%
\bibitem [{\citenamefont {Sau}\ \emph {et~al.}(2010{\natexlab{b}})\citenamefont
  {Sau}, \citenamefont {Tewari}, \citenamefont {Lutchyn}, \citenamefont
  {Stanescu},\ and\ \citenamefont {Das~Sarma}}]{Sau}%
  \BibitemOpen
  \bibfield  {author} {\bibinfo {author} {\bibfnamefont {J.~D.}\ \bibnamefont
  {Sau}}, \bibinfo {author} {\bibfnamefont {S.}~\bibnamefont {Tewari}},
  \bibinfo {author} {\bibfnamefont {R.~M.}\ \bibnamefont {Lutchyn}}, \bibinfo
  {author} {\bibfnamefont {T.~D.}\ \bibnamefont {Stanescu}}, \ and\ \bibinfo
  {author} {\bibfnamefont {S.}~\bibnamefont {Das~Sarma}},\ }\bibfield  {title}
  {\enquote {\bibinfo {title} {Non-abelian quantum order in spin-orbit-coupled
  semiconductors: Search for topological majorana particles in solid-state
  systems},}\ }\href@noop {} {\bibfield  {journal} {\bibinfo  {journal} {Phys.
  Rev. B}\ }\textbf {\bibinfo {volume} {82}},\ \bibinfo {pages} {214509}
  (\bibinfo {year} {2010}{\natexlab{b}})}\BibitemShut {NoStop}%
\bibitem [{\citenamefont {Duckheim}\ and\ \citenamefont
  {Brouwer}(2011)}]{Brouwer2011}%
  \BibitemOpen
  \bibfield  {author} {\bibinfo {author} {\bibfnamefont {M.}~\bibnamefont
  {Duckheim}}\ and\ \bibinfo {author} {\bibfnamefont {P.~W.}\ \bibnamefont
  {Brouwer}},\ }\bibfield  {title} {\enquote {\bibinfo {title} {Andreev
  reflection from noncentrosymmetric superconductors and majorana bound-state
  generation in half-metallic ferromagnets},}\ }\href {\doibase
  10.1103/PhysRevB.83.054513} {\bibfield  {journal} {\bibinfo  {journal} {Phys.
  Rev. B}\ }\textbf {\bibinfo {volume} {83}},\ \bibinfo {pages} {054513}
  (\bibinfo {year} {2011})}\BibitemShut {NoStop}%
\bibitem [{\citenamefont {Chung}\ \emph {et~al.}(2011)\citenamefont {Chung},
  \citenamefont {Zhang}, \citenamefont {Qi},\ and\ \citenamefont
  {Zhang}}]{ZhangPRB2011}%
  \BibitemOpen
  \bibfield  {author} {\bibinfo {author} {\bibfnamefont {S.~B.}\ \bibnamefont
  {Chung}}, \bibinfo {author} {\bibfnamefont {H.-J.}\ \bibnamefont {Zhang}},
  \bibinfo {author} {\bibfnamefont {X.-L.}\ \bibnamefont {Qi}}, \ and\ \bibinfo
  {author} {\bibfnamefont {S.-C.}\ \bibnamefont {Zhang}},\ }\bibfield  {title}
  {\enquote {\bibinfo {title} {Topological superconducting phase and majorana
  fermions in half-metal/superconductor heterostructures},}\ }\href {\doibase
  10.1103/PhysRevB.84.060510} {\bibfield  {journal} {\bibinfo  {journal} {Phys.
  Rev. B}\ }\textbf {\bibinfo {volume} {84}},\ \bibinfo {pages} {060510}
  (\bibinfo {year} {2011})}\BibitemShut {NoStop}%
\bibitem [{\citenamefont {Potter}\ and\ \citenamefont
  {Lee}(2012)}]{SOCTsc_PotterLee_2012}%
  \BibitemOpen
  \bibfield  {author} {\bibinfo {author} {\bibfnamefont {A.~C.}\ \bibnamefont
  {Potter}}\ and\ \bibinfo {author} {\bibfnamefont {P.~A.}\ \bibnamefont
  {Lee}},\ }\bibfield  {title} {\enquote {\bibinfo {title} {Topological
  superconductivity and majorana fermions in metallic surface states},}\ }\href
  {\doibase 10.1103/PhysRevB.85.094516} {\bibfield  {journal} {\bibinfo
  {journal} {Phys. Rev. B}\ }\textbf {\bibinfo {volume} {85}},\ \bibinfo
  {pages} {094516} (\bibinfo {year} {2012})}\BibitemShut {NoStop}%
\bibitem [{\citenamefont {R\"ontynen}\ and\ \citenamefont
  {Ojanen}(2015)}]{SOCchiralp_OjanenPRL2015}%
  \BibitemOpen
  \bibfield  {author} {\bibinfo {author} {\bibfnamefont {J.}~\bibnamefont
  {R\"ontynen}}\ and\ \bibinfo {author} {\bibfnamefont {T.}~\bibnamefont
  {Ojanen}},\ }\bibfield  {title} {\enquote {\bibinfo {title} {Topological
  superconductivity and high chern numbers in 2d ferromagnetic shiba
  lattices},}\ }\href {\doibase 10.1103/PhysRevLett.114.236803} {\bibfield
  {journal} {\bibinfo  {journal} {Phys. Rev. Lett.}\ }\textbf {\bibinfo
  {volume} {114}},\ \bibinfo {pages} {236803} (\bibinfo {year}
  {2015})}\BibitemShut {NoStop}%
\bibitem [{\citenamefont {Li}\ \emph {et~al.}(2016)\citenamefont {Li},
  \citenamefont {Neupert}, \citenamefont {Wang}, \citenamefont {MacDonald},
  \citenamefont {Yazdani},\ and\ \citenamefont {Bernevig}}]{Li:2016aa}%
  \BibitemOpen
  \bibfield  {author} {\bibinfo {author} {\bibfnamefont {J.}~\bibnamefont
  {Li}}, \bibinfo {author} {\bibfnamefont {T.}~\bibnamefont {Neupert}},
  \bibinfo {author} {\bibfnamefont {Z.}~\bibnamefont {Wang}}, \bibinfo {author}
  {\bibfnamefont {A.~H.}\ \bibnamefont {MacDonald}}, \bibinfo {author}
  {\bibfnamefont {A.}~\bibnamefont {Yazdani}}, \ and\ \bibinfo {author}
  {\bibfnamefont {B.~A.}\ \bibnamefont {Bernevig}},\ }\bibfield  {title}
  {\enquote {\bibinfo {title} {Two-dimensional chiral topological
  superconductivity in shiba lattices},}\ }\href {\doibase 10.1038/ncomms12297}
  {\bibfield  {journal} {\bibinfo  {journal} {Nat. Commun.}\ }\textbf {\bibinfo
  {volume} {7}},\ \bibinfo {pages} {12297} (\bibinfo {year}
  {2016})}\BibitemShut {NoStop}%
\bibitem [{\citenamefont {Hsu}\ \emph {et~al.}(2017)\citenamefont {Hsu},
  \citenamefont {Vaezi}, \citenamefont {Fischer},\ and\ \citenamefont
  {Kim}}]{Hsu2017}%
  \BibitemOpen
  \bibfield  {author} {\bibinfo {author} {\bibfnamefont {Y.-T.}\ \bibnamefont
  {Hsu}}, \bibinfo {author} {\bibfnamefont {A.}~\bibnamefont {Vaezi}}, \bibinfo
  {author} {\bibfnamefont {M.~H.}\ \bibnamefont {Fischer}}, \ and\ \bibinfo
  {author} {\bibfnamefont {E.-A.}\ \bibnamefont {Kim}},\ }\bibfield  {title}
  {\enquote {\bibinfo {title} {Topological superconductivity in monolayer
  transition metal dichalcogenides},}\ }\href {\doibase 10.1038/ncomms14985}
  {\bibfield  {journal} {\bibinfo  {journal} {Nat. Commun.}\ }\textbf {\bibinfo
  {volume} {8}},\ \bibinfo {pages} {14985} (\bibinfo {year}
  {2017})}\BibitemShut {NoStop}%
\bibitem [{\citenamefont {Wang}\ \emph {et~al.}(2022)\citenamefont {Wang},
  \citenamefont {Powers}, \citenamefont {Zhang}, \citenamefont {Smith},
  \citenamefont {McIntosh}, \citenamefont {Bac}, \citenamefont {Riney},
  \citenamefont {Zhukovskyi}, \citenamefont {Orlova}, \citenamefont
  {Rokhinson}, \citenamefont {Hsu}, \citenamefont {Liu},\ and\ \citenamefont
  {Assaf}}]{Wang_NanoLett2022}%
  \BibitemOpen
  \bibfield  {author} {\bibinfo {author} {\bibfnamefont {J.}~\bibnamefont
  {Wang}}, \bibinfo {author} {\bibfnamefont {W.}~\bibnamefont {Powers}},
  \bibinfo {author} {\bibfnamefont {Z.}~\bibnamefont {Zhang}}, \bibinfo
  {author} {\bibfnamefont {M.}~\bibnamefont {Smith}}, \bibinfo {author}
  {\bibfnamefont {B.~J.}\ \bibnamefont {McIntosh}}, \bibinfo {author}
  {\bibfnamefont {S.~K.}\ \bibnamefont {Bac}}, \bibinfo {author} {\bibfnamefont
  {L.}~\bibnamefont {Riney}}, \bibinfo {author} {\bibfnamefont
  {M.}~\bibnamefont {Zhukovskyi}}, \bibinfo {author} {\bibfnamefont
  {T.}~\bibnamefont {Orlova}}, \bibinfo {author} {\bibfnamefont {L.~P.}\
  \bibnamefont {Rokhinson}}, \bibinfo {author} {\bibfnamefont {Y.-T.}\
  \bibnamefont {Hsu}}, \bibinfo {author} {\bibfnamefont {X.}~\bibnamefont
  {Liu}}, \ and\ \bibinfo {author} {\bibfnamefont {B.~A.}\ \bibnamefont
  {Assaf}},\ }\bibfield  {title} {\enquote {\bibinfo {title} {Observation of
  coexisting weak localization and superconducting fluctuations in strained
  sn1--xinxte thin films},}\ }\bibfield  {booktitle} {\emph {\bibinfo
  {booktitle} {Nano Letters}},\ }\href {\doibase 10.1021/acs.nanolett.1c04370}
  {\bibfield  {journal} {\bibinfo  {journal} {Nano Lett.}\ }\textbf {\bibinfo
  {volume} {22}},\ \bibinfo {pages} {792--800} (\bibinfo {year}
  {2022})}\BibitemShut {NoStop}%
\bibitem [{\citenamefont {Jang}\ \emph {et~al.}(2017)\citenamefont {Jang},
  \citenamefont {Yoo}, \citenamefont {Pfeiffer}, \citenamefont {West},
  \citenamefont {Baldwin},\ and\ \citenamefont {Ashoori}}]{MERTS}%
  \BibitemOpen
  \bibfield  {author} {\bibinfo {author} {\bibfnamefont {J.}~\bibnamefont
  {Jang}}, \bibinfo {author} {\bibfnamefont {H.~M.}\ \bibnamefont {Yoo}},
  \bibinfo {author} {\bibfnamefont {L.~N.}\ \bibnamefont {Pfeiffer}}, \bibinfo
  {author} {\bibfnamefont {K.~W.}\ \bibnamefont {West}}, \bibinfo {author}
  {\bibfnamefont {K.~W.}\ \bibnamefont {Baldwin}}, \ and\ \bibinfo {author}
  {\bibfnamefont {R.~C.}\ \bibnamefont {Ashoori}},\ }\bibfield  {title}
  {\enquote {\bibinfo {title} {Full momentum- and energy-resolved spectral
  function of a 2d electronic system},}\ }\href {\doibase
  10.1126/science.aam7073} {\bibfield  {journal} {\bibinfo  {journal}
  {Science}\ }\textbf {\bibinfo {volume} {358}},\ \bibinfo {pages} {901--906}
  (\bibinfo {year} {2017})}\BibitemShut {NoStop}%
\bibitem [{\citenamefont {Huang}\ \emph {et~al.}(2023)\citenamefont {Huang},
  \citenamefont {Yue}, \citenamefont {Baydin}, \citenamefont {Zhu},
  \citenamefont {Nojiri}, \citenamefont {Kono}, \citenamefont {He},\ and\
  \citenamefont {Yi}}]{ARPES_field}%
  \BibitemOpen
  \bibfield  {author} {\bibinfo {author} {\bibfnamefont {J.}~\bibnamefont
  {Huang}}, \bibinfo {author} {\bibfnamefont {Z.}~\bibnamefont {Yue}}, \bibinfo
  {author} {\bibfnamefont {A.}~\bibnamefont {Baydin}}, \bibinfo {author}
  {\bibfnamefont {H.}~\bibnamefont {Zhu}}, \bibinfo {author} {\bibfnamefont
  {H.}~\bibnamefont {Nojiri}}, \bibinfo {author} {\bibfnamefont
  {J.}~\bibnamefont {Kono}}, \bibinfo {author} {\bibfnamefont {Y.}~\bibnamefont
  {He}}, \ and\ \bibinfo {author} {\bibfnamefont {M.}~\bibnamefont {Yi}},\
  }\bibfield  {title} {\enquote {\bibinfo {title} {{Angle-resolved
  photoemission spectroscopy with an in situ tunable magnetic field}},}\
  }\href@noop {} {\bibfield  {journal} {\bibinfo  {journal} {Review of
  Scientific Instruments}\ }\textbf {\bibinfo {volume} {94}},\ \bibinfo {pages}
  {093902} (\bibinfo {year} {2023})}\BibitemShut {NoStop}%
\bibitem [{\citenamefont {Dong}\ \emph {et~al.}(2020)\citenamefont {Dong},
  \citenamefont {Xiao}, \citenamefont {Xiong},\ and\ \citenamefont
  {Niu}}]{Niu2020}%
  \BibitemOpen
  \bibfield  {author} {\bibinfo {author} {\bibfnamefont {L.}~\bibnamefont
  {Dong}}, \bibinfo {author} {\bibfnamefont {C.}~\bibnamefont {Xiao}}, \bibinfo
  {author} {\bibfnamefont {B.}~\bibnamefont {Xiong}}, \ and\ \bibinfo {author}
  {\bibfnamefont {Q.}~\bibnamefont {Niu}},\ }\bibfield  {title} {\enquote
  {\bibinfo {title} {Berry phase effects in dipole density and the mott
  relation},}\ }\href {\doibase 10.1103/PhysRevLett.124.066601} {\bibfield
  {journal} {\bibinfo  {journal} {Phys. Rev. Lett.}\ }\textbf {\bibinfo
  {volume} {124}},\ \bibinfo {pages} {066601} (\bibinfo {year}
  {2020})}\BibitemShut {NoStop}%
\bibitem [{\citenamefont {Parameswaran}\ \emph {et~al.}(2012)\citenamefont
  {Parameswaran}, \citenamefont {Kivelson}, \citenamefont {Shankar},
  \citenamefont {Sondhi},\ and\ \citenamefont {Spivak}}]{Spivak}%
  \BibitemOpen
  \bibfield  {author} {\bibinfo {author} {\bibfnamefont {S.~A.}\ \bibnamefont
  {Parameswaran}}, \bibinfo {author} {\bibfnamefont {S.~A.}\ \bibnamefont
  {Kivelson}}, \bibinfo {author} {\bibfnamefont {R.}~\bibnamefont {Shankar}},
  \bibinfo {author} {\bibfnamefont {S.~L.}\ \bibnamefont {Sondhi}}, \ and\
  \bibinfo {author} {\bibfnamefont {B.~Z.}\ \bibnamefont {Spivak}},\ }\bibfield
   {title} {\enquote {\bibinfo {title} {Microscopic model of quasiparticle wave
  packets in superfluids, superconductors, and paired hall states},}\
  }\href@noop {} {\bibfield  {journal} {\bibinfo  {journal} {Phys. Rev. Lett.}\
  }\textbf {\bibinfo {volume} {109}},\ \bibinfo {pages} {237004} (\bibinfo
  {year} {2012})}\BibitemShut {NoStop}%
\bibitem [{FN0()}]{FN0}%
  \BibitemOpen
  \href@noop {} {}\bibinfo {note} {Here we keep $\partial_{\kb} \chi$ for a
  general pairing symmetry, but later we will restrict ourselves to the case of
  $s$-wave with $\partial_{\kb} \chi=0$ for the evaluation of BC correction to
  observables.}\BibitemShut {Stop}%
\bibitem [{\citenamefont {Culcer}\ \emph {et~al.}(2005)\citenamefont {Culcer},
  \citenamefont {Yao},\ and\ \citenamefont {Niu}}]{NiuMult}%
  \BibitemOpen
  \bibfield  {author} {\bibinfo {author} {\bibfnamefont {D.}~\bibnamefont
  {Culcer}}, \bibinfo {author} {\bibfnamefont {Y.}~\bibnamefont {Yao}}, \ and\
  \bibinfo {author} {\bibfnamefont {Q.}~\bibnamefont {Niu}},\ }\bibfield
  {title} {\enquote {\bibinfo {title} {Coherent wave-packet evolution in
  coupled bands},}\ }\href@noop {} {\bibfield  {journal} {\bibinfo  {journal}
  {Phys. Rev. B}\ }\textbf {\bibinfo {volume} {72}},\ \bibinfo {pages} {085110}
  (\bibinfo {year} {2005})}\BibitemShut {NoStop}%
\bibitem [{FN1()}]{FN1}%
  \BibitemOpen
  \href@noop {} {}\bibinfo {note} {The wave-packet average of an operator
  $\hat{O}$ refers to $ \,\bra{W_s} \hat{O} \ket{W_s} - \bra{G} \hat{O}
  \ket{G}$.}\BibitemShut {Stop}%
\bibitem [{LL()}]{LL}%
  \BibitemOpen
  \href@noop {} {}\bibinfo {note} {L. Landau, E. Lifshitz, Electrodynamics of
  Continuous Media (Pergamon, New York, 1960).}\BibitemShut {Stop}%
\bibitem [{Arn()}]{Arnold}%
  \BibitemOpen
  \href@noop {} {}\bibinfo {note} {V.I. Arnold, Mathematical Methods of
  Classical Mechanics (Springer, New York, 1978)}\BibitemShut {NoStop}%
\bibitem [{Sch()}]{Schrieffer}%
  \BibitemOpen
  \href@noop {} {}\bibinfo {note} {J. R. Schrieffer, Theory of
  Superconductivity (Perseus Books, Reading, MA, 1983).}\BibitemShut {Stop}%
\bibitem [{FN2()}]{FN2}%
  \BibitemOpen
  \href@noop {} {}\bibinfo {note} {The contribution from the penetration depth
  is expected to be negligible in the weak field regime, where the vortices are
  sparse compared to the beam size.}\BibitemShut {Stop}%
\end{thebibliography}%
		
\end{document}